\documentclass[sigconf, nonacm]{acmart}

\usepackage{algorithmic}
\usepackage{float}
\usepackage{stfloats}
\usepackage{graphicx}
\usepackage{subcaption}
\usepackage{textcomp}
\usepackage[dvipsnames]{xcolor}
\usepackage{multirow}
\usepackage{times}
\usepackage[normalem]{ulem}
\usepackage{url}
\usepackage{lipsum}
\usepackage[utf8]{inputenc}
\usepackage{caption}
\usepackage{amsmath}
\usepackage{amsthm}
\usepackage{booktabs}
\usepackage[switch]{lineno}
\usepackage{kotex}
\usepackage{threeparttable}
\usepackage[linesnumbered,algoruled,boxed,ruled,noend]{algorithm2e}
\usepackage{hyperref}
\usepackage{verbatim}
\usepackage{xcolor,colortbl}
\usepackage{float}
\usepackage{lineno}

\newcommand{\kgp}{{$(k,g,p)$-core}}
\newcommand{\kg}{{$(k,g)$-core}}
\newcommand{\naive}{\textsf{NPA}}
\newcommand{\npa}{\textsf{NPA}}
\newcommand{\ASAP}{\textsf{ASAP}}
\newcommand{\asap}{\textsf{ASAP}}
\newcommand{\NPA}{\textsf{NPA}}

\newcommand{\spara}[1]{\smallskip\noindent{\bf #1}}

\usepackage{amsmath}

\newtheorem{definition}{Definition}
\newtheorem{property}{Property}

\newtheorem{example}{Example}

\DeclareRobustCommand{\hourglass}
{\mathrel{\mathpalette\hour@glass\relax}}
\usepackage{enumitem}

\def\boxit#1#2#3{
  \setlength{\fboxsep}{5.4pt}
  \setlength{\fboxrule}{1pt}
  \makebox[0pt][l]{
    \color{blue}
    \hspace{#1}
    \raisebox{#3}{
      \fbox{\makebox[#2][c]{}}
    }
  }
}

\begin{document}

\title{Beyond Trivial Edges: A Fractional Approach to Cohesive Subgraph Detection in Hypergraphs}


\author{Hyewon Kim}
\affiliation{%
  \institution{Ulsan National Institute of\\Science and Technology}
  \city{}\country{}}
\email{hyewon.kim@unist.ac.kr}

\author{Woocheol Shin}
\affiliation{%
  \institution{Ulsan National Institute of\\Science and Technology}
  \city{}\country{}}
\email{woofe@unist.ac.kr}

\author{Dahee Kim}
\affiliation{%
  \institution{Ulsan National Institute of\\Science and Technology}
  \city{}\country{}}
\email{dahee@unist.ac.kr}

\author{Junghoon Kim}
\affiliation{%
  \institution{Ulsan National Institute of\\Science and Technology}
  \city{}\country{}}
\email{junghoon.kim@unist.ac.kr}
\authornote{Corresponding author}

\author{Sungsu Lim}
\affiliation{%
  \institution{Chungnam National University}
  \city{}\country{}}
\email{sungsu@cnu.ac.kr}

\author{Hyun Ji Jeong}
\affiliation{%
  \institution{Kongju National University}
  \city{}\country{}}
\email{hjjeong@kongju.ac.kr}

\renewcommand{\shortauthors}{H. Kim et al.}

\begin{abstract}
  Hypergraphs can capture high-order relationships in complex systems, yet large hyperedges often dilute cohesive structures by incorporating loosely related nodes. To address this, we propose a \emph{fraction-based cohesive subgraph} model, called the $(k,g,p)$-core, which extends existing support-based frameworks by introducing a user-defined fraction threshold. This threshold effectively filters out hyperedges deemed too large to convey meaningful connections, thereby emphasising high-quality, context-specific relationships. We devise two algorithms—\emph{Naïve} and \emph{Advanced}—to efficiently compute the $(k,g,p)$-core. The Advanced algorithm leverages lazy update strategies to avoid repeated neighbour recalculations, reducing computational overhead. Experimental evaluations on real-world datasets show that our method not only preserves the accuracy of cohesive subhypergraph discovery but also improves computational efficiency by over $50\%$ compared to baseline approaches. Our findings demonstrate the importance of fraction-based constraints in refining subhypergraph discovery, opening avenues for more robust hypergraph analysis in domains such as recommendation systems, anomaly detection, and community detection.
\end{abstract}





\maketitle

\section{INTRODUCTION}\label{sec:introduction}
Graphs have become key tools across various domains of network analysis, providing fundamental insights into the connectivity, dynamics, and structure of complex systems~\cite{easley2010networks}. From social network analysis, where they map interactions among individuals or organisations, to biological networks that explain the intricate pathways of genes and proteins~\cite{yang2020graph}, graphs serve as the backbone for understanding the relationships and dependencies that manage these complex systems~\cite{barabasi2013network}. They enable researchers to analyse interactions at various levels, facilitating the discovery of patterns, the prediction of behaviours, and the identification of key nodes in networks~\cite{das2018study}. 

With the capability to address various types of data and the continuous advancement of technology, there has recently been a focus on studying, exploring, and managing various types of graphs~\cite{chakrabarti2006graph,camacho2020four}. Among the various types of graphs, hypergraphs have gained significant attention due to their ability to capture high-order relationships beyond pairwise interactions~\cite{bretto2013hypergraph}, thereby allowing them to be suitable for many applications~\cite{bick2023higher}. A hypergraph is a more general type of graph that allows edges, known as hyperedges, to connect several nodes simultaneously~\cite{berge1984hypergraphs}. This feature is particularly effective in representing complex and high-order relationships that cannot be reduced to simple pairwise interactions~\cite{10598068}. 

For instance, in co-authorship networks, a conventional graph structure limits each edge to connect only two researchers~\cite{roy2015measuring}. This framework becomes problematic when trying to represent papers co-authored by three or more researchers, as it fails to accurately reflect the equal level of collaboration among all contributors. In contrast, hypergraphs allow for more efficient and accurate modelling of these complex associations by enabling a single hyperedge to connect multiple researchers simultaneously, thus naturally representing group collaborations more effectively.

Due to their structure where hyperedges can connect more than two nodes, hypergraphs possess distinct characteristics compared to conventional graphs: (1) \textit{Nested hyperedge structure}~\cite{lotito2022higher}: Hyperedges can be nested within each other, introducing a hierarchical structure and containment properties. This characteristic allows hypergraphs to represent complex relationships that involve grouping or hierarchical organisation of hyperedges. 
(2) \textit{Heavy-tailed distribution}~\cite{lee2024survey}: The distributions of degrees, the cardinalities, and intersection sizes of hyperedges follow a heavy-tailed distribution. Thus, most nodes have a relatively low degree, but a few have a very high degree. Similarly, most hyperedges have a low cardinality, with only a few having a very high cardinality. The intersection sizes also show this distribution. This pattern is indicative of scale-free networks and is observed in many real-world networks.
These unique properties have made them a powerful tool for modelling complex systems and relationships.

Consequently, the research of hypergraph mining has recently been conducted in areas such as recommendation systems~\cite{xia2021self}, anomaly detection~\cite{liang2021industrial}, frequent pattern mining~\cite{ranshous2017exchange}, contagion dynamics analysis~\cite{luo2024hierarchical},  bioinformatics~\cite{feng2021hypergraph}, transportation systems~\cite{wang2021metro}, and text mining~\cite{bazaga2024hyperbert}. 
Due to its utility in various applications~\cite{10598134}, the study of cohesive subhypergraph discovery~\cite{leng2013m, bu2023hypercore, nbrkcore, batagelj2011fast, alphabeta, kgcore, alphabetap, kpcore, kim2024experimental} has obtained considerable attention. The concept of a cohesive subhypergraph is a set of nodes that are densely connected by hyperedges. This structure enables the representation of high-order and complex interactions that cannot be captured by traditional graphs. Thus, it is important for understanding the underlying structure of complex systems, and for uncovering deeper insights into these networks.

Even though there are numerous studies on cohesive subgraph discovery in traditional networks~\cite{malliaros2020core}, these approaches cannot be directly applied to hypergraphs due to their unique characteristics.
Therefore, research on the discovery of various cohesive subhypergraphs, including the Clique-core~\cite{batagelj2011fast}, the $k$-hypercore~\cite{k-hypercore}, and others, have been also proposed and are detailed in Section~\ref{sec:relatedwork}. Among them, the $(k,g)$-core~\cite{kgcore} is recently proposed by extending the core concept~\cite{seidman1983network} by incorporating both the number of the neighbours and its support values, allowing it to capture the high-level cohesiveness of the subhypergraph. 
However, the $(k,g)$-core has a critical limitation: it fails to filter out very large hyperedges that encompass loosely connected or irrelevant nodes.
Since $(k,g)$-core relies solely on neighbour size and support values, large hyperedges unintentionally inflate the support values of weakly connected nodes, causing them to be retained in subgraphs despite lacking strong interactions. This leads to misleading cohesive subgraphs, where nodes appear significant even though they only participate in trivial hyperedges. Thus, a more refined approach is required to suppress the influence of such hyperedges while preserving genuinely cohesive structures.

\begin{figure}[h]
    \includegraphics[width=0.99\linewidth]{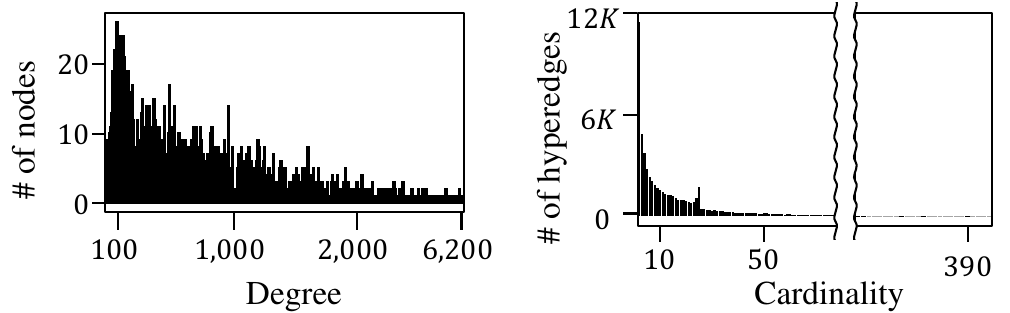}
    \caption{\mbox{Degree and cardinality distribution}}
    \label{fig:intro_dist}
\end{figure}

This limitation is evident across real-world datasets~\cite{Benson2018subset,chodrow2021hypergraph,Fowler-2006-connecting,Fowler-2006-cosponsorship,kunegis2013konect,mayer2018hype}, where large hyperedges distort analytical results. Figure~\ref{fig:intro_dist} shows the degree and cardinality distribution for the House Bills dataset. Most hyperedges contain fewer than $50$ nodes, but some hyperedges, denoted as trivial hyperedges, exhibit exceptionally high cardinalities, which skew the substructure. Notably, the largest hyperedge in the House Bills contains $26.7\%$ of all nodes, while the top $10$ hyperedges with the highest cardinalities combined cover $58.97\%$. These trivial hyperedges disproportionately influence connectivity patterns, leading to overestimated support values that distort cohesive subgraphs and obscure meaningful communities. As a result, nodes with small degrees, which form the majority, are influenced by these trivial hyperedges, leading to distorted interpretations of their connectivity and structural importance.

The $(k,g)$-core relies on neighbour size and support values, with parameter selection often guided by degree distributions to identify cohesive subhypergraphs. However, trivial hyperedges inflate the degrees of weakly connected nodes, distorting connectivity patterns and cohesion metrics. This leads to misleading conclusions and may obscure community structures or fail to identify key nodes within hypergraphs.
To illustrate these challenges and underscore the need for refining existing cohesive models, we present a motivating example below.

\begin{figure}[t]
\includegraphics[width=0.999\linewidth]{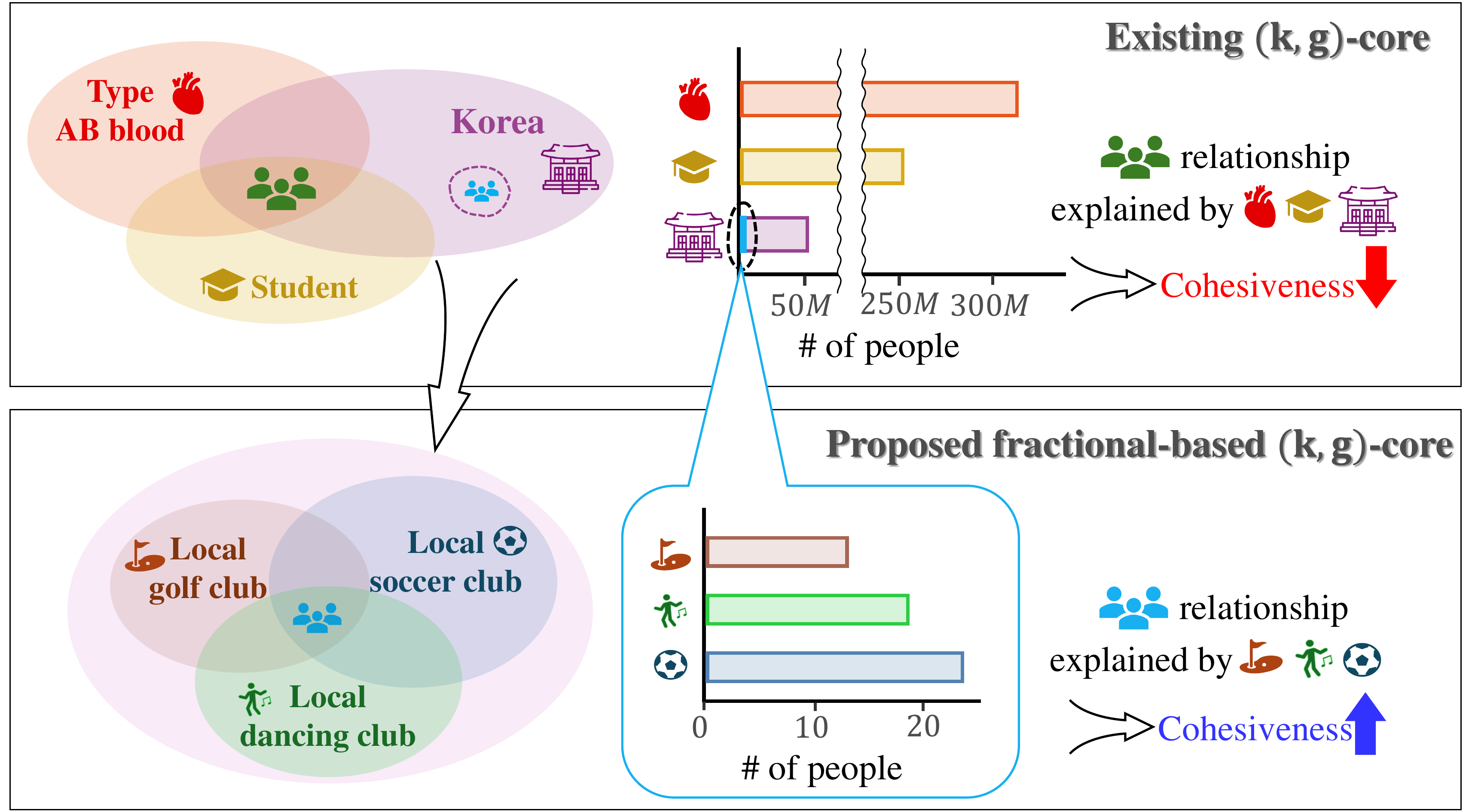}
    \caption{Motivating example}
    \label{fig:intro}
\end{figure}

\begin{example}\label{example:1}
In Figure~\ref{fig:intro}, the user group (green colour) on the top side has the same blood type, is composed of students, and lives in Korea. Similarly, the user group (blue colour) on the bottom side belongs to the same local Korea golf club, soccer club, and dancing club. For the green group, the connections are overly broad and may fail to capture a strong bond of the user group, i.e., the relationships explaining their cohesion are too ``trivial''. Conversely, the blue group on the bottom side shares ``local'' level small-sized groups, making their relationships relatively narrower and easier to explain. This implies that to find a blue-coloured group, knowing they live in Korea alone is insufficient to capture the true cohesion of the group. In this work, we aim to intentionally avoid considering trivial relationships to find cohesive user groups. 
\end{example}

To handle the limitations of the $(k,g)$-core, we propose a novel cohesive subhypergraph model that effectively de-emphasises relatively less important hyperedges. 
Determining which hyperedges are trivial is an open question and challenge; thus, we define our model based on the intuition that the size of hyperedges affects the cohesion of the user group. As shown in Example~\ref{example:1}, the more general the concept, the more nodes it can include, which can be considered insufficient to model cohesiveness. To address this, we present a user parameter that defines how meaningful a hypergraph is based on its size. This parameter allows us to disregard relatively less important hyperedges, thereby enabling the discovery of more meaningful and cohesive subhypergraphs.

Therefore, in this paper, we propose a new model, the $(k,g,p)$-core, which extends the $(k,g)$-core by incorporating a fraction threshold for hyperedges to address its limitations (Section~\ref{sec:kgp_intro}). This model effectively retrieves cohesive subhypergraphs by ensuring that the subhypergraphs consist of sufficiently close ($p$), frequently co-occurring ($g$), and connected ($k$) neighbours. To find the $(k,g,p)$-core, we present two peeling algorithms: 
\begin{itemize}[leftmargin=*]
    \item Na\"ive Peeling Algorithm (\NPA): This algorithm iteratively removes hyperedges and nodes that violate the $(k,g,p)$ constraints. However, it recomputes neighbour support values from scratch at each iteration, making it computationally expensive.
    \item Advanced Support-based Algorithm with Pruning (\ASAP): To mitigate inefficiency, {\ASAP} introduces two key optimisations: (i) a supporting table that tracks changes in node connectivity, allowing for lazy updates instead of full recomputations, and (ii) pruning strategies that use node- and edge-based lower bounds to reduce unnecessary neighbour recalculations. These optimisations reduce redundant operations, significantly improving computational efficiency.
\end{itemize}

\spara{Applications.} The applications of our model are as follows.
\begin{enumerate}[leftmargin=*]
    \item \textit{\underline{E-commerce product recommendation}:} The $(k,g,p)$-core can significantly enhance product recommendation systems~\cite{wang2022hypergraph} in e-commerce platforms. In an e-commerce environment, customers and products form a hypergraph, where a hyperedge represents a customer's purchase of multiple products. One key advantage of using the $(k,g,p)$-core enables the filtering of customers who have purchased an excessively large number of products. This prevents recommendations from being dominated by these outliers and ensures that the resulting subgroups consist of customers with more typical purchasing patterns.
    \item \textit{\underline{Frequent pattern mining}:} Frequent pattern mining~\cite{luna2019frequent} is a key problem in data mining. However, traditional methods often impose strict frequency requirements~\cite{gan2017mining,min2020frequent}, which can lead to the exclusion of significant patterns that appear frequently but not uniformly. The $(k,g,p)$-core provides a more flexible approach by incorporating the fraction parameter $p$. This allows for finding cohesive subhypergraphs that capture frequently occurring patterns, even if they do not satisfy the strict frequency thresholds of traditional methods. By leveraging the fraction parameter, the $(k,g,p)$-core represents more meaningful and varied patterns within large and complex datasets.
    \item \mbox{\textit{\underline{A key step for other hypergraph problems}:} Computing the} $(k,g,p)$-core can serve as an essential step in addressing other significant hypergraph problems, such as hypergraph clustering~\cite{mayer2018hype} and community detection~\cite{ma2023hypergraph,chien2018community}. By identifying cohesive subhypergraphs, the $(k,g,p)$-core provides a foundational step in tackling other problems, which are essential in various applications such as network analysis~\cite{young2021hypergraph}.
\end{enumerate}

\spara{Contributions.} Three major contributions of this paper are summarised as follows. 
\begin{itemize}[leftmargin=*]
    \item \textit{\underline{{Novel Fraction-Based Model}}:} We propose the $(k,g,p)$-core, introducing a fraction threshold to mitigate the adverse impact of trivial hyperedges in hypergraphs.
    \item \textit{\underline{Algorithmic Framework}:} We design two peeling algorithms, Naïve (\NPA) and Advanced (\ASAP), that demonstrate how pruning strategies and lazy updates can significantly reduce the time required to discover cohesive subhypergraphs.
    \item \textit{\underline{Comprehensive Evaluation}:} Through extensive experiments on real-world and synthetic datasets, we demonstrate that our approach outperforms existing solutions in both runtime and quality, achieving over a $50\%$ improvement in computational efficiency.
\end{itemize}

\spara{Reproducibility.} We publish our code at \url{https://github.com/hwhwkim7/kgpcore.git}

\spara{Paper structure.} The structure of this paper is organised as follows: Section~\ref{sec:kgp_intro} presents the proposed $(k,g,p)$-core model. Section~\ref{sec:kgcore_comp} presents the peeling algorithm for $(k,g)$-core computation. Section~\ref{sec:kgp_alg} proposes two algorithms with pruning strategies to find $(k,g,p)$-core. Section~\ref{sec:select_p} discusses a reuse strategy designed to reduce computational overhead significantly. Section~\ref{sec:experiments} demonstrates the performance of our algorithms through various experiments in real-world datasets. Section~\ref{sec:relatedwork} discusses the related cohesive subgraph models. Finally, Section~\ref{sec:conclusion} concludes our work and presents the future directions. 

\section{PROBLEM STATEMENT}\label{sec:kgp_intro}
This section introduces the {\kgp} problem, including key definitions and a fraction constraint in hypergraphs.

\subsection{Preliminaries}
In this work, we consider an unweighted and undirected hypergraph. The formal definition is as follows.

\begin{definition}\label{def:hypergraph}
(\underline{Hypergraph}). A hypergraph is a graph $G = (V, E)$, where:  $V$ is a finite set of nodes; and $E$ is a set of hyperedges, where each hyperedge $e \in E$ is a non-empty subset of $V$, i.e., $e \subseteq V$.
\end{definition}

In contrast to traditional graphs, where edges connect exactly two nodes, a hyperedge can connect any number of nodes. Given a subset $H \subseteq V$, we define $G[H]$ as the induced subhypergraph by $H \subseteq V$ with its hyperedges given by $E[H] = \{e | \text{node}(e) \cap H \neq \emptyset\}$. Further notations are summarised in Table~\ref{tab:notations}. We next define two basic concepts that are used in hypergraph analysis. 

\begin{definition}
(\underline{Cardinality}). Given a hypergraph $G=(V,E)$, a set of nodes $H \subseteq V$, and a hyperedge $e\in E$, the cardinality of a hyperedge $e$ is the number of nodes in $e$ that are in $H$.
\end{definition}

\begin{definition}
(\underline{Degree}). Given a hypergraph $G=(V,E)$, a set of hyperedges $E' \subseteq E$, and a node $v \in V$, the degree of a node $v$ is the number of hyperedges in $E'$ that contain $v$.
\end{definition}

In this work, we aim to identify cohesive subhypergraphs in order to understand their complicated relationships and patterns. The following are two key definitions:

\begin{table}[t]
\small
\centering
\caption{Notations}
\label{tab:notations}
\begin{tabular}{c||l}
\hline
\textbf{Notation} & \textbf{Description} \\ \hline \hline
$G$ & a hypergraph, $G=(V,E)$ \\ \hline
node$(e)$ & a set of nodes in hyperedge $e$ in $G$ \\ \hline
edge$(v)$ & a set of hyperedges which contain a node $v$ in $G$ \\ \hline
$f(e,H)$ & a fraction of $e$ in $H$, i.e., $|\text{node}(e)\cap H|/|\text{node}(e)|$ \\ \hline 
$sup(v,u,E)$ & the number of hyperedges containing $u$ and $v$ in $E$ \\ \hline 
$N^g(v)$ & the set of $g$-neighbours of $v$\\ \hline 
$S[v]$ & the number $g$-neighbours of $v$ \\ \hline
$M[v]$ & the supporting table of $v$ \\ \hline
\hline
\end{tabular}
\end{table}

\begin{definition}\label{def:sup}
(\underline{Support}). Given a hypergraph $G=(V,E)$, the support value of a node $v \in V$ with a node $u \in V \setminus \{v\}$, denoted as $sup(v, u, E)$, is the number of hyperedges that contain both nodes $v$ and $u$ in $E$.
\end{definition}

\begin{definition}\label{def:g_neighbours}
(\underline{$g$-neighbours}). Given a hypergraph $G=(V,E)$ and an integer $g$, the $g$-neighbours of a node $v \in V$ are the set of all nodes $u \in V \setminus \{v\}$ such that $sup(v, u, E) \geq g$.
\end{definition}

In the context of social networks, $g$-neighbours can be conceptualised as ``close friends'' who share multiple group affiliations or interests with node $v$, indicating a strong social bond. 
In transaction records, they represent ``frequently co-purchased items'' that appear together in at least $g$ times, indicating a strong associative relationship. 
With these concepts, the $(k,g)$-core~\cite{kgcore} is defined. 

\begin{definition}\label{def:kgcore}
    (\underline{$(k,g)$-core}~\cite{kgcore}). Given a hypergraph $G=(V,E)$, $k\ge 1$, and $g\ge 1$, the $(k,g)$-core is a maximal node set in which each node has at least $k$ neighbours which appear in at least $g$ hyperedges together. 
\end{definition}

The $(k,g)$-core exhibits a hierarchical structure with respect to both $k$ and $g$, and for any fixed pair of parameters $(k,g)$, the resulting subhypergraph is unique. However, large hyperedges can artificially inflate support values and cause weakly related nodes to remain in the structure, leading to distorted or misleading cohesive subhypergraphs. 

\begin{figure*}[t]
    \begin{subfigure}{.55\linewidth}
\includegraphics[width=0.99\linewidth]{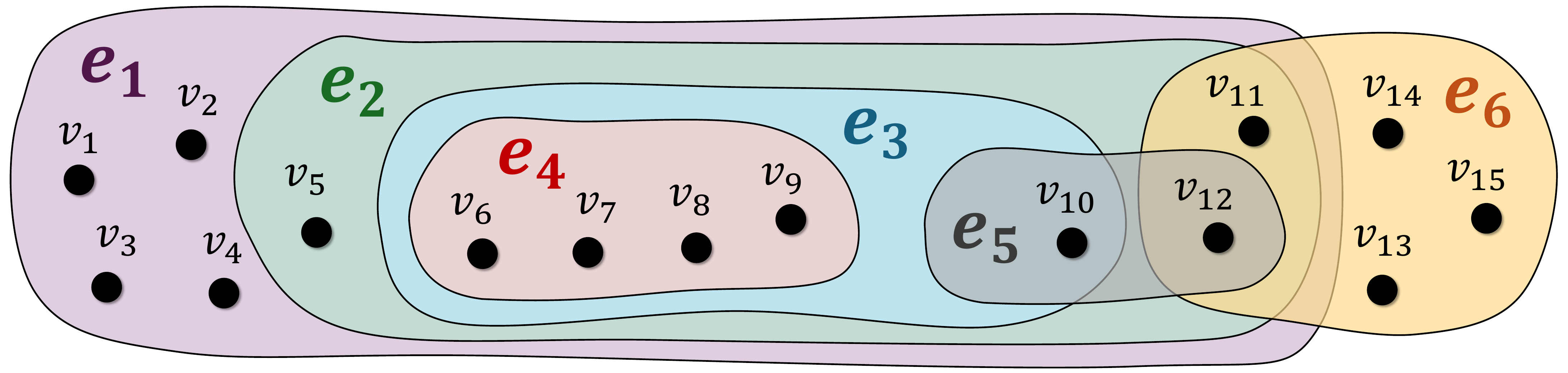}
    \caption{Hypergraph}
    \label{fig:toy_graph_hyp}
    \end{subfigure}
    \begin{subfigure}{.44\linewidth}
\includegraphics[width=0.99\linewidth]{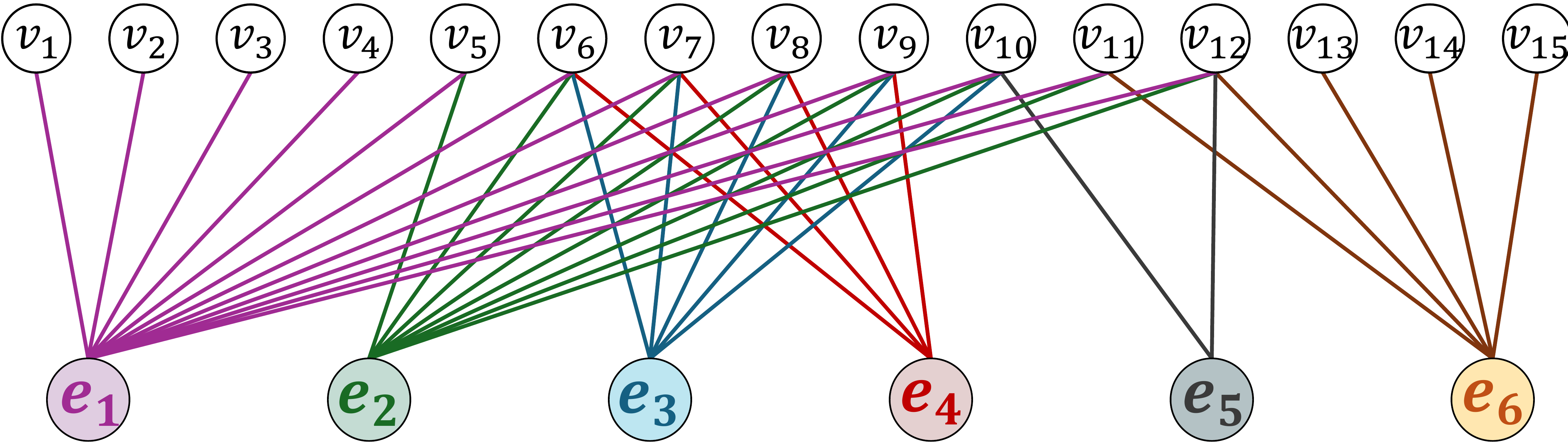}
    \caption{Corresponding bipartite graph}
    \label{fig:toy_graph_bip}
    \end{subfigure}
    \caption{A simple example}
    \label{fig:toy_graph}
\end{figure*}

\begin{example}
    Figure~\ref{fig:toy_graph_hyp} shows a hypergraph $G=(V,E)$ with $15$ nodes and $6$ hyperedges, and Figure~\ref{fig:toy_graph_bip} represents the corresponding graph converted to a bipartite graph. Given $k=2$ and $g=2$, the nodes $\{v_1, \cdots , v_4, v_{13}, \cdots, v_{15}\}$ are removed, because they only have $1$-neighbours. The remaining nodes each have two or more size of $2$-neighbours, so nodes $\{v_5, \cdots, v_{12}\}$ belong to $(2,2)$-core. 
\end{example}

\subsection{Problem definition}
In this section, we introduce the $(k,g,p)$-core and its characteristics. We begin by presenting the concept of a fraction, which distinguishes this model from the $(k,g)$-core.

\begin{definition}\label{definition:fraction}
    (\underline{Fraction}). Given a hypergraph $G=(V,E)$, a set of nodes $H \subseteq V$, and a hyperedge $e \in E$, the fraction of a hyperedge $e$ is the ratio of the number of nodes in $e$ that are also in $H$ to the total number of nodes in $e$, i.e., $f(e, H) = |\text{node}(e) \cap H| / |\text{node}(e)|$.
\end{definition}

According to Definition~\ref{definition:fraction}, the range of the fraction is $0 \le f(e, H) \le 1$. $f(e, H) = 0$ indicates that a hyperedge $e$ does not contain any nodes in $H$. Conversely, $f(e, H) = 1$ implies that all nodes of hyperedge $e$ are in $H$.

Utilising a fraction threshold allows us to classify hyperedges within a hypergraph into two categories: (1) \textit{weak hyperedge} and (2) \textit{strong hyperedge}. Given a set of nodes $H$, a hyperedge $e$, and a fraction threshold $p$, we categorise a hyperedge $e$ as a \textit{weak hyperedge} if the fraction of nodes in $e$ that are also in $H$ is less than $p$. Conversely, if this fraction is at least $p$, then $e$ is classified as a \textit{strong hyperedge}. The formal definitions are as follows:

\begin{definition}\label{definition:weak_hyperedge}
    (\underline{Weak hyperedge}). Given a hypergraph $G=(V,E)$, a fraction threshold $p$, and a set of nodes $H \subseteq V$, a hyperedge $e \in E$ is a weak hyperedge if and only if $f(e,H) < p$.
\end{definition}

\begin{definition}\label{definition:strong_hyperedge}
    (\underline{Strong hyperedge}). Given a hypergraph $G=(V,E)$, a fraction threshold $p$, and a set of nodes $H \subseteq V$, a hyperedge $e \in E$ is a strong hyperedge if and only if $f(e,H) \geq p$.
\end{definition}

According to Definitions~\ref{definition:weak_hyperedge} and~\ref{definition:strong_hyperedge}, we propose a new cohesive model, the $(k,g,p)$-core, which extends the $(k,g)$-core~\cite{kgcore} by incorporating a fraction threshold for hyperedges to address the limitations, i.e., this model requires sufficient strong hyperedges to form a cohesive subhypergraph. 
The formal definition of the $(k,g,p)$-core is as follows.

\begin{definition}\label{definition:kgt}
    \mbox{(\underline{$(k,g,p)$-core}). Given a hypergraph $G=(V,E)$,} $k \geq 1$, $g \geq 1$, and $p \in [0,1]$, the $(k,g,p)$-core, denoted as $H$, is a set of nodes in $V$ that satisfies the following constraints:
    \begin{itemize}[leftmargin=*]
        \item $H$ is a maximal set of nodes.
        \item For every node $v \in H$, there exist at least $k$ neighbours of $v$ within $H$ that, together with $v$, appear in at least $g$ strong hyperedges, determined by $p$, in $G[H]$.
    \end{itemize}
\end{definition}

The $(k,g,p)$-core model extends the traditional $(k,g)$-core model, incorporating an additional parameter $p$. This extension retains two properties of the $(k,g)$-core: uniqueness and hierarchical structure, which provide deeper insights into the structural composition of $(k,g,p)$-core.

\begin{property}
    Given a hypergraph $G=(V, E)$ with specific $k,g$ and $p$, the $(k,g,p)$-core is unique.
\end{property}

\begin{proof}
Suppose that we have two distinct $(k,g,p)$-cores, $H_1$ and $H_2$, such that $H_1 \neq H_2$. Consider the union of two sets $H_{12} = H_1 \cup H_2$. The set  $H_{12}$ intrinsically satisfies the constraints for the $(k,g,p)$-core since the fraction of involved hyperedges must be increased. This contradicts the maximality constraint, as the size of $H_{12}$ must be larger than either $H_1$ or $H_2$. It implies that $H_1$ or $H_2$ does not satisfy the maximality constraint. Thus, the assumption that $H_1 \neq H_2$ is not true, meaning that the $(k,g,p)$-core is unique. 
\end{proof}

\begin{property}    
    Given a hypergraph $G=(V,E)$ with specific $k,g$ and $p$, the $(k,g,p)$-core has hierarchical structure.
\end{property}

\begin{proof}
Based on the $(k,g)$-core~\cite{kgcore}, the hierarchical structure is already proven. Thus, we prove that the $(k,g,p)$-core also has hierarchical structure with a fraction threshold $p$. Assume that for $p' \ge p$ and fixed values $k$ and $g$, a node $v$ belongs to the $(k,g,p')$-core but not to the $(k,g,p)$-core. As the value of $p$ increases, it imposes a stricter constraint, leading to the removal of more hyperedges and consequently returning a smaller node set. Consequently, contrary to the initial assumption, if $v$ belongs to the $(k,g,p')$-core, it must also belong to the $(k,g,p)$-core. Thus, $(k,g,p')$-core $\subseteq (k,g,p)$-core if $p' \ge p$ and $p, p' \in [0,1]$.
\end{proof}

\begin{figure}[t]
\includegraphics[width=0.99\linewidth]{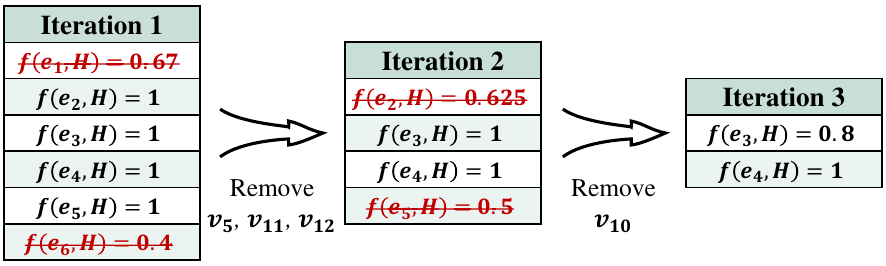}
    \caption{The fraction computation of Figure~\ref{fig:toy_graph_hyp} with $p=0.7$}
    \label{fig:fraction_computation}
\end{figure}

\begin{example}\label{example:intro_example} 
Given a hypergraph in Figure~\ref{fig:toy_graph_hyp}, $k=2$, $g=2$, and $p=0.7$, the $(2,2)$-core is a set of nodes $H = \{v_5, \cdots, v_{12}\}$. The fraction computation of the hyperedges $e\in E$ with $H$ is as illustrated in Figure~\ref{fig:fraction_computation}. Note that hyperedges $e_1$ and $e_6$ have the fractions below the fraction threshold $p$. This leads to the removal of nodes $v_5$, $v_{11}$, and $v_{12}$, due to violations of the $k$ and $g$ constraints. The removal of these nodes changes the fractions of other hyperedges; as a result, hyperedges $e_2$ and $e_5$ are also removed. Through iterative application of these criteria, the final $(2,2,0.7)$-core is determined to be $\{v_6, \cdots, v_9\}$.
\end{example}

\section{$(k,g)$-CORE COMPUTATION}\label{sec:kgcore_comp}

\begin{algorithm}[!b]
\footnotesize
\SetAlgoLined
\SetKwData{break}{break}
\SetKwData{AND}{AND}
\SetKwData{false}{false}
\SetKwData{true}{true}
\SetKwData{del}{delete}
\SetKwFunction{queueInit}{queue}
\SetKwFunction{emptyc}{empty}
\SetKwFunction{pop}{pop}
\SetKwFunction{contain}{contain}
\SetKwFunction{cnt}{s}
\SetKwFunction{push}{push}
\SetKwFunction{getOccurMap}{getOccurMap}
\SetKwFunction{getNbrMap}{getNbrMap}
\SetKwFunction{getKeys}{getKeys}
\KwIn{Hypergraph $G=(V,E)$, parameters $k$ and $g$}
\KwOut{\mbox{$V'$: The $(k, g)$-core of $G$, $S$: $g$-neighbours size}}
$S \leftarrow \{\}$\;
$V' \leftarrow V$\;
$VQ \leftarrow$ \queueInit{}\;
\For{$v \in V'$}{
    $N^g (v) \leftarrow$ \getNbrMap{$V', E', v, g$}\;
    $S[v] \leftarrow \left|N^g (v)\right|$\;
    \If{$S[v] < k$}{
        $VQ$.\push{$v$}\;
    }
}
\While{$VQ$.\emptyc{} $\neq$ \true}{
    $v \leftarrow VQ$.\pop{}\; 
    $N^g (v) \leftarrow$ \getNbrMap{$V', E', v, g$}\;
    $V' \leftarrow V' \setminus \{v\}$\;
    \del $S[v]$\;
    \For{$w \in$ \getKeys{$N^g (v)$}}{
        \If{$VQ$.\contain{$w$} $\neq$ \true}{
            $S[w] \leftarrow S[w]-1$\;
            \If{$S[w]<k$}{
                $VQ$.\push{$w$}\;
            }
        }
    }
}
\Return{$V'$, $S$}
\caption{Efficient $(k,g)$-core computation}
\label{alg:kgcore_algorithm}
\end{algorithm}

In this section, we present an algorithm to find the $(k,g)$-core in hypergraphs. Compared to the peeling algorithm~\cite{kgcore}, we develop a space-efficient peeling algorithm without requiring extensive memory space while preserving the same time complexity. In \cite{kgcore}, the authors describe a straightforward approach to compute the $(k,g)$-core, which maintains a set of neighbours and their support values for every node in the hypergraph. This requires $O(|V|^2)$ space as a node may have up to $|V|-1$ neighbours. However, the presence of large trivial hyperedges inherently demands a significant amount of memory space. To resolve this issue, we present a new algorithm that maintains the time complexity while avoiding excessive memory consumption and ensuring efficiency in handling large datasets. 

\spara{Algorithmic procedure.} 
The procedure for computing the $(k,g)$-core employs a peeling algorithm that iteratively eliminates nodes failing to satisfy the constraints until no removable nodes remain. The function \FuncSty{getNbrMap} is defined as $\{(u,  \text{sup}(v,u,E')) |u \in V'$, sup$(v,u,E') \geq g\}$. This approach significantly reduces memory consumption by maintaining only the count of $g$-neighbours for each node, thereby determining its eligibility for inclusion in the $(k,g)$-core. The detailed pseudo-description is provided in Algorithm~\ref{alg:kgcore_algorithm}.

\spara{Theoretical analysis.} 
The time complexity of Algorithm~\ref{alg:kgcore_algorithm} is $O(|V|^2|E|)$. The number of iterations needed to remove a set of nodes is $O(|V|)$, and the process of checking whether the neighbours of each node satisfy the support constraint takes $O(|V||E|)$. Our algorithm requires only $O(|V|)$ additional memory to store the count of $g$-neighbours for each node, making it more efficient than storing a neighbour list in memory~\cite{kgcore}, which demands $O(|V|^2)$ space.

\section{$(k,g,p)$-CORE COMPUTATION}\label{sec:kgp_alg}
In this section, we present two algorithms: the Na\"ive Peeling Algorithm (\NPA) and the Advanced Support-based Algorithm with Pruning strategies (\ASAP). The {\NPA} serves as a baseline algorithm with a straightforward approach, primarily focusing on the direct computation of the core structures. In contrast, the {\ASAP} algorithm improves efficiency by employing effective pruning strategies that prevent inefficient recomputations and thereby reduce overall computational overhead.

\subsection{\underline{N}a\"ive \underline{P}eeling \underline{A}lgorithm (\naive)}
We first present the Na\"ive peeling algorithm that iteratively removes hyperedges and nodes until the remaining subhypergraph satisfies the constraints.

\begin{algorithm}[h]
\footnotesize
\SetAlgoLined
\SetKwData{break}{break}
\SetKwData{AND}{AND}
\SetKwData{false}{false}
\SetKwData{true}{true}
\SetKwData{del}{delete}
\SetKwFunction{queueInit}{queue}
\SetKwFunction{emptyc}{empty}
\SetKwFunction{pop}{pop}
\SetKwFunction{contain}{contain}
\SetKwFunction{push}{push}
\SetKwFunction{getOccurMap}{getOccurMap}
\SetKwFunction{size}{size}
\SetKwFunction{kgcore}{kgcore}
\SetKwFunction{node}{node}
\SetKwFunction{edge}{edge}
\SetKwFunction{clear}{clear}
\SetKwFunction{getNbrMap}{getNbrMap}
\SetKwFunction{getKeys}{getKeys}
\SetKwRepeat{Do}{do}{while}
\KwIn{Hypergraph $G=(V,E)$, parameters $k$, $g$ and $p$}
\KwOut{The $(k, g, p)$-core of $G$}
$VQ \leftarrow$ \queueInit{}\;
$VC \leftarrow \{\}$\; 
$V', S \leftarrow$ \kgcore{$G,k,g$}\;
$E', EC \leftarrow$ $\{e | \text{\node{$e$}} \cap V' \neq \emptyset \}, \forall e\in E$\;
\Do{$EC$.\emptyc{} $\neq$ \true}{
    \For{$e \in EC$}{
        \If{$f(e, V') < p$}{
            $VC \leftarrow VC \cup  \{\text{\node{$e$}} \cap V'$\}\;
            $E' \leftarrow E' \setminus \{e\}$\;
        }
    }
    \For{$v \in VC$}{
        $N^g (v) \leftarrow$ \getNbrMap{$V', E', v, g$}\;
        $S[v] \leftarrow |N^g (v)|$\;
        \If{$S[v] < k$}{
            $VQ$.\push{$v$}\;
        }
    }
    $EC$.\clear{}, $VC$.\clear{}\;
    \While{$VQ$.\emptyc{} $\neq$ \true}{
        $v \leftarrow VQ$.\pop{}\; 
        $EC \leftarrow EC \cup  \{\text{\edge{$v$}} \cap E'$\}\;
        $N^g (v) \leftarrow$ \getNbrMap{$V', E', v, g$}\;
        $V' \leftarrow V' \setminus \{v\}$\;
        \del $S[v]$\;
        \For{$w \in $\getKeys{$N^g (v)$}}{
            \If{$VQ$.\contain{$w$} $\neq$ \true}{
                $S[w] \leftarrow S[w]-1$\;
                \If{$S[w]<k$}{
                    $VQ$.\push{$w$}\;
                }
            }
        }
    }
}
\Return{$V'$}
\caption{Procedure of the {\npa}}
\label{alg:kgpcore_naive}
\end{algorithm}

\spara{Algorithmic procedure.} 
The {\NPA} has two procedures: (1) hyperedge peeling step; and (2) node peeling step. Algorithm~\ref{alg:kgpcore_naive} provides a pseudo-description of the {\NPA}. The hyperedge peeling step involves iteratively removing hyperedges whose fractions do not satisfy the specified threshold $p$ (Lines 6-9). Subsequently, the node peeling step is applied, identifying nodes that no longer satisfy the $k$ and $g$ constraints (Lines 10-14). Then, it iteratively removes nodes in a cascading manner and finds a candidate set of hyperedges for the removal process (Lines 16-26). This process is repeated until no more hyperedges or nodes remain for removal.

\spara{Theoretical analysis.}
Note that computing $N^g(v)$ for any node $v$ requires $O(|V||E|)$ time. In the worst-case scenario, a loop that iteratively removes nodes requires up to $O(|V|)$ times, thereby contributing to a time complexity of $O(|V|^2|E|)$. We denote the total number of iterations of the entire procedure by $t$, which is bounded by $\min(|V|,|E|)$. Consequently, the time complexity of the {\NPA} is $O(t|V|^2|E|)$.

\begin{example}
Considering Figure~\ref{fig:toy_graph_hyp} with parameters $k=2$, $g=2$, and $p=0.7$, the {\NPA} procedure is as follows: As an initial step, it obtains a set of nodes $V' = \{v_5, \cdots, v_{12}\}$ and computes the count of $g$-neighbours for each node, stored in $S$. In the first iteration, hyperedges $e_1$ and $e_6$ are removed as their fractions are $0.67$ and $0.4$, respectively. Subsequently, all nodes from $e_1$ and $e_6$ are placed into the candidate node set $VC$, and the $g$-neighbours sizes are updated such that $S[v_6] =\cdots= S[v_9] = 4$, $S[v_{10}] = 5$, $S[v_5] =  S[v_{11}] = 0$ and $S[v_{12}] =1$. Next, nodes $v_5$, $v_{11}$, and $v_{12}$, such that $S[v]$ is less than $k$, are removed. Consequently, hyperedges $e_2$ and $e_5$ are added to the candidate removal hyperedge set $EC$. This process is repeated until the {\NPA} finally returns $V' = \{v_6, \cdots, v_9\}$.
\end{example}

\begin{figure}[t]
\includegraphics[width=0.99\linewidth]{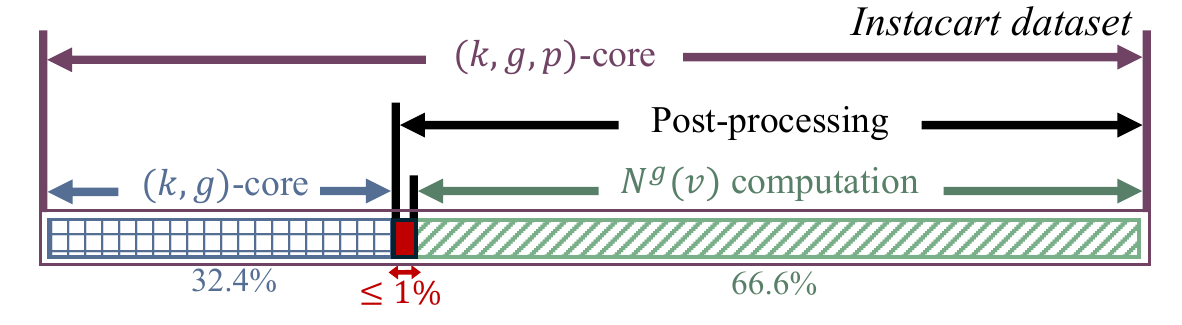}
    \vspace{-0.2cm}
    \caption{Component-based execution time in the {\NPA}}
    \vspace{-0.2cm}
    \label{fig:asap_motivation}
\end{figure}

\spara{Limitation of the {\NPA}.}
A key concern of the {\NPA} is the repeated computation of $N^g(v)$, which has a time complexity of $O(|V||E|)$. Figure~\ref{fig:asap_motivation} shows the component-based execution time on the Instacart dataset~\cite{Benson2018subset}. We divide the $(k,g,p)$-core computation process into the $(k,g)$-core computation and a post-processing procedure. We notice that the computation of $N^g(v)$ takes nearly $67\%$ of the total running time. The high cost of computing $N^g(v)$, which needs to be run many times, indicates potential for optimisation. To reduce the frequency of $N^g(v)$ computations, we manage sketched information about $g$-neighbours for each node, specifically maintaining a count of nodes for each support value. Thus, we employ a lazy update strategy with lower-bounds to efficiently avoid unnecessary recomputations. Applying this strategy on the Instacart dataset reduces the $N^g(v)$ computation frequency by approximately $56\%$ (from $35,953$ to $15,571$).

\subsection{\underline{A}dvanced \underline{S}upport-based \underline{A}lgorithm with \underline{P}runing strategies (\asap)}
We present an Advanced Support-based Algorithm with Pruning strategies, called {\ASAP}, which extends the {\NPA} by incorporating pruning strategies. Its primary goal is to reduce repeated computations of $g$-neighbours, which is considered a dominant operation in the {\NPA} due to its prohibitive time complexity ($O(|V||E|)$). 

\subsubsection{\bf{Preliminaries}}
We introduce a compact data structure  called a supporting table. Every node maintains its own supporting table, and it consists of the number of nodes corresponding to each support value.

\begin{definition}
(\underline{Supporting table}.)
Given a hypergraph $G=(V, E)$ with a support threshold $g$, a supporting table $M[v]$ for a node $v \in V$ is a dictionary. 
\begin{itemize}[leftmargin=*]
\item Each key in this dictionary represents a unique \textit{support} value associated with node $v$.
\item For each key, its value denotes the number of neighbours of node $v$ whose support value is equal to the key. 
\end{itemize}
Note that support values in $M$ must be greater than or equal to the support threshold $g$.
\end{definition}

For every node, the number of rows of its supporting table is different. Notably, the sum of the values for each node in the supporting table indicates its minimum number of $g$-neighbours. 

\spara{Space analysis.} The maximum size of the supporting table $M[v]$ of node $v$ is bounded by the minimum of the number of hyperedges containing $v$ and the size of its $g$-neighbours, i.e., $O(\min(|e(v)|, |N^g(v)|))$.

\begin{figure}[h]
    \includegraphics[width=0.99\linewidth]{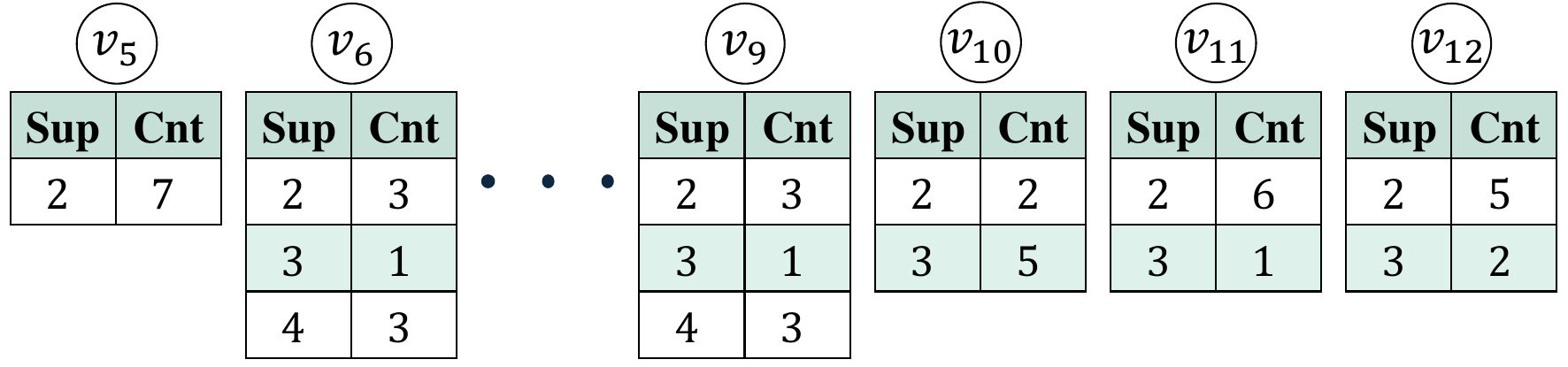}
    \caption{A supporting table of Figure~\ref{fig:toy_graph_hyp}}
    \vspace{-0.2cm}
    \label{fig:table}
\end{figure}

\begin{example}
    In Figure~\ref{fig:toy_graph_hyp}, which presents a hypergraph with parameters $k=2$ and $g=2$, the resultant supporting table is illustrated in Figure~\ref{fig:table}. For node $v_6$, the supporting table has a support value $2$ with nodes $\{v_5, v_{11}, v_{12}\}$ through hyperedges $\{e_1, e_2\}$, a support value $3$ with node $\{v_{10}\}$ through hyperedges $\{e_1, e_2, e_3\}$, and a support value $4$ with nodes $\{v_7, v_8, v_9\}$ through hyperedges $\{e_1, e_2, e_3, e_4\}$.
\end{example}

\subsubsection{\bf{Supporting table based pruning strategies}}

\begin{figure*}[t]
\begin{subfigure}{.33\linewidth}
\includegraphics[width=0.99\linewidth]{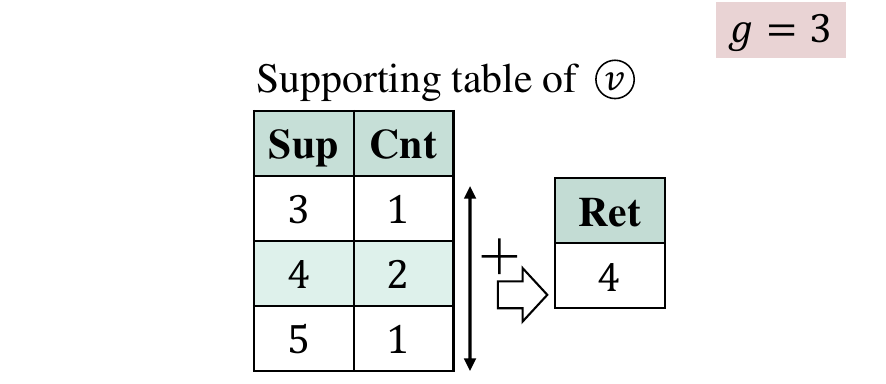}
\caption{Getting $k$ from supporting table}
\label{fig:ub}
\end{subfigure}
\begin{subfigure}{.33\linewidth}
\includegraphics[width=0.99\linewidth]{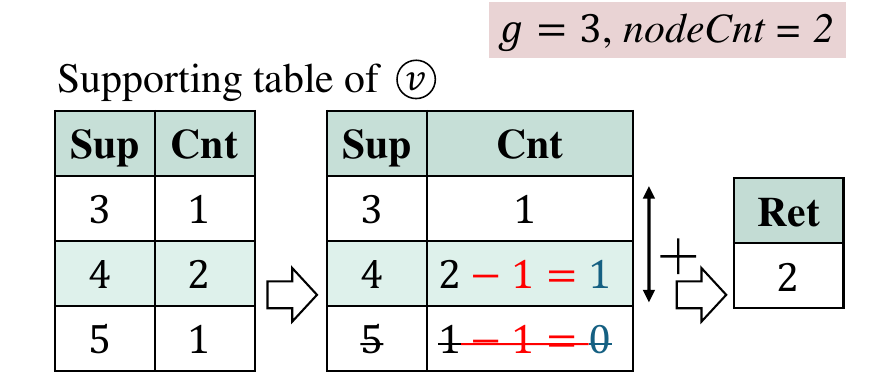}
\caption{Getting lower-bound with $nodeCnt$}
\label{fig:nodeUB}
\end{subfigure}
\begin{subfigure}{.33\linewidth}
\includegraphics[width=0.99\linewidth]{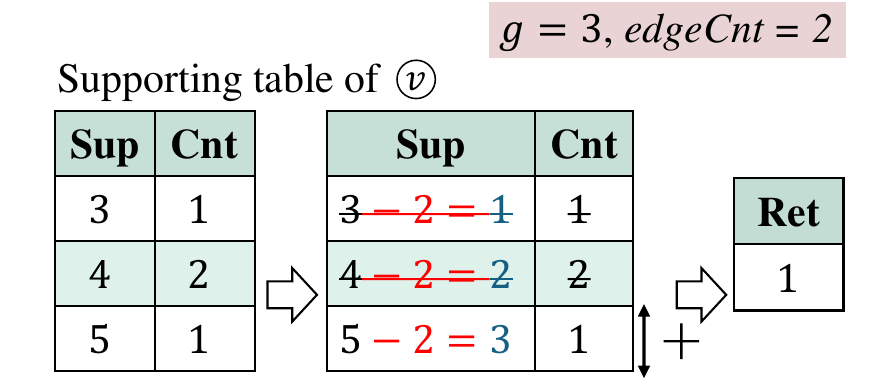}
\caption{Getting lower-bound with $edgeCnt$}
\label{fig:edgeUB}
\end{subfigure}
\caption{Computing lower-bounds}
\vspace{-0.3cm}
\label{fig:lower_bound_example}
\end{figure*}

This section presents key pruning techniques to reduce repeated computations of $g$-neighbours using the supporting tables. These tables are updated infrequently in a lazy manner based on predetermined lower-bounds, which significantly enhances computational efficiency.

\begin{algorithm}[h]
\footnotesize
\SetAlgoLined
\SetKwData{sumV}{sum}
\SetKwFunction{emptyc}{empty}
\SetKwFunction{getKeys}{getKeys}
\SetKwFunction{get}{get}
\KwIn{Supporting table $M[v]$}
\KwOut{Minimum $g$-neighbour size}
$M\leftarrow M[v]$\;
\If{$M$.\emptyc{}}{
    \Return 0\;
}
\sumV $\leftarrow 0$\;
\For{$y \in $ \getKeys{$M$}}{
    $\sumV \leftarrow \sumV + M.$\get{$y$}\;
}
\Return \sumV\;
\caption{\mbox{\FuncSty{getEstNbrSize()}: Getting estimated } $g$-neighbours size from a supporting table}
\label{alg:est_gnbr}
\end{algorithm}

\spara{Estimating $g$-neighbours size.} Given a node $v$ and its corresponding supporting table $M[v]$, we estimate the number of the $g$-neighbours by summing all values in $M[v]$. Since the keys of the table are support values greater than or equal to $g$, the sum indicates the size of $g$-neighbours illustrated in Figure~\ref{fig:ub}. Noted that the derived $g$-neighbours count may be less than or equal to the actual $g$-neighbours size due to the lazy update approach. The details are in the Algorithm~\ref{alg:est_gnbr}.

\begin{algorithm}[h]
\footnotesize
\SetAlgoLined
\SetKwData{sumV}{sum}
\SetKwData{del}{del}
\SetKwData{null}{null}
\SetKw{break}{break}
\SetKwFunction{getK}{getEstNbrSize}
\SetKwFunction{remove}{remove}
\SetKwFunction{get}{get}
\SetKwFunction{decrease}{decreasedgeCnt}
\SetKwFunction{putEntry}{put}
\SetKwFunction{getMaxKey}{getMaxSup}
\KwIn{A node $v$, supporting table $M[v]$}
\KwOut{Minimum $g$-neighbour size}
$c \leftarrow v.nodeCnt$\;
$M\leftarrow M[v]$\;
\While{$c \neq 0$}{
    $mk \leftarrow $ \getMaxKey{$M$}\;
    \If{$mk = $ \null}{
        \Return 0\;
    }
    \If{$c \ge$ $M.$\get{$mk$}}{
        $c \leftarrow c-M.$\get{$mk$}\;
        $M.$\remove{$mk$}\;
    }
    \Else{
        $M.$\putEntry{$mk, M.$\get{$mk$}$-c$}\;
        \break\;
    }
}
$v.nodeCnt \leftarrow 0$\;
\Return \getK{$M$}\;
\caption{\mbox{\FuncSty{getNodeLB()}: Node-based lower-bound}}
\label{alg:lb_nodeCnt}
\end{algorithm}

\spara{Estimating $g$-neighbours size with the number of removed $g$-neighbours.}
Given a node $v$, its supporting table $M[v]$, and the number of removed $g$-neighbours (denoted as $nodeCnt$), we propose a method to estimate the number of $g$-neighbours using the supporting table. In this approach, we consider the worst-case scenario by removing the node with the highest support value from the table first. This is because the node with the maximum support value, representing the most stable and closely connected neighbours, may still not be removed after several hyperedges are processed. Therefore, by removing it, we can obtain the lower-bound of $g$-neighbours under extreme conditions. The detailed procedures are described in Algorithm~\ref{alg:lb_nodeCnt} and Figure~\ref{fig:nodeUB}. Note that a supporting table which is only updated based on the $nodeCnt$, always returns the exact size of $g$-neighbours.

\begin{algorithm}[h]
\footnotesize
\SetAlgoLined
\SetKw{Return}{return}
\SetKw{break}{break}
\SetKwFunction{sortByKeys}{sortByKeys}
\SetKwFunction{putEntry}{put}
\SetKwFunction{get}{get}
\SetKwFunction{remove}{remove}
\SetKwFunction{getKeys}{getKeys}
\SetKwFunction{contain}{contain}
\SetKwFunction{entry}{getEntry}
\KwIn{A node $v$, supporting table $M[v]$, parameter $g$}
\KwOut{Minimum $g$-neighbour size}
$c \leftarrow v.edgeCnt$\;
$M\leftarrow M[v]$\;
$T \leftarrow \{\}$\;
\ForEach{$(s, cnt)$ in $M$.\entry{}}{
    \If{$s - c \ge g$}{
        $T.$\putEntry{$s-c, cnt$}\;
    }
}
$v.edgeCnt \leftarrow 0$\;
$M \leftarrow T$\;
\Return \getK{$M$}\;
\caption{\mbox{\FuncSty{getEdgeLB()}: Edge-based lower-bound}}
\label{alg:lb_edgeCnt}
\end{algorithm}

\spara{Estimating $g$-neighbours size with the number of removed hyperedges.}
Given a node $v$, its supporting table $M[v]$, and the number of removed hyperedges containing $v$ (denoted as $edgeCnt$), we propose a method to estimate the size of $g$-neighbours using the supporting table. In this approach, we consider the worst-case scenario where all $g$-neighbours of $v$ are contained within all removed hyperedges, leading to a decrease in all support values by $edgeCnt$. Note that obtaining a tight lower-bound on the size of $g$-neighbours is inherently challenging due to the complexity of hypergraph structures. Therefore, rather than attempting to derive a tight lower-bound, our method provides an estimate under extreme conditions to offer a conservative assessment of the $g$-neighbours size. This approach provides a practical estimation even in the worst-case scenario. However, more precise and refined techniques for estimating $g$-neighbours sizes merit deeper investigation in future work. The details are in Algorithm~\ref{alg:lb_edgeCnt} and Figure~\ref{fig:edgeUB}.

\begin{algorithm}[!b]
\footnotesize
\SetAlgoLined
\SetKwData{break}{break}
\SetKwData{AND}{AND}
\SetKwData{false}{false}
\SetKwData{true}{true}
\SetKwData{del}{delete}
\SetKwFunction{queueInit}{queue}
\SetKwFunction{emptyc}{empty}
\SetKwFunction{pop}{pop}
\SetKwFunction{contain}{contain}
\SetKwFunction{cnt}{s}
\SetKwFunction{push}{push}
\SetKwFunction{add}{add}
\SetKwFunction{getEdgeLB}{getEdgeLB}
\SetKwFunction{getLB}{getLB}
\SetKwFunction{getNodeLB}{getNodeLB}
\SetKwFunction{initialise}{initialisedgeCnt}
\SetKwFunction{getNbrMap}{getNbrMap}
\SetKwFunction{size}{size}
\SetKwFunction{kgcore}{kgcore}
\SetKwFunction{node}{node}
\SetKwFunction{update}{updateSupTable}
\SetKwFunction{edge}{edge}
\SetKwFunction{clear}{clear}
\SetKwFunction{getKeys}{getKeys}
\SetKwFunction{insert}{insert}
\SetKwFunction{len}{len}
\SetKwFunction{pqInit}{priorityqueue}
\SetKwFunction{IncnodeCnt}{IncreasenodeCnt}
\SetKwFunction{valueSum}{valSum}
\SetKwFunction{max}{max}
\SetKwRepeat{Do}{do}{while}
\SetKwBlock{BlockIf}{\mbox{\FuncSty{/* Lines 19-32 in Algorithm~\ref{alg:asap_alg} */}}}{}
\SetKwBlock{BlockFor}{\llap  ...}

\KwIn{Hypergraph $G=(V,E)$, parameters $k$, $g$ and $p$}
\KwOut{The $(k, g, p)$-core of $G$}
$VC \leftarrow \{\}$\;
$V', S \leftarrow$ \kgcore{$G,k,g$}\;
$PQ \leftarrow$ \pqInit{}\;\
$E', EC \leftarrow$ $\{e | \text{\node{$e$}} \cap V' \neq \emptyset \}, \forall e\in E$\;
$M \leftarrow \{\}$\;
\For{$v\in$ \getKeys{$S$}}{
    $M[v].$\insert{$g$, $S[v]$}\;
}
\Do{$EC$.\emptyc{} $\neq$ \true}{
    \For{$e \in EC$}{
        \If{$f(e, V') < p$}{
            \If{$|\text{\node{$e$}} \cap V'| > 1$}{
                \For{$v \in$ \text{\node{$e$}} $\cap$ $V'$}{
                    $v.edgeCnt\leftarrow v.edgeCnt+1$\;
                    $VC.$\add{$v$}\;
                }
            }
            $E' \leftarrow E' \setminus \{e\}$\;
        }
    }
    $EC$.\clear{}\;
    \For{$v \in VC$}{
        \If{\getEdgeLB{$v$, $M[v]$, $g$} $< k$}{ 
            $N^g (v) \leftarrow$ \getNbrMap{$V', E', v, g$}\;
            \If{$\left|N^g(v) \right| < k$}{
                $EC \leftarrow EC \cup \{\text{\edge{$v$}} \cap E'$\}\;
                $V' \leftarrow V' \setminus \{v\}$\; 
                $M.$\remove{$v$}\;
                \If{$PQ$.\contain{$v$}}{
                    $PQ$.\remove{$v$}\;
                }
                \For{$w \in$ \getKeys{$N^g(v)$}}{
                    $w.nodeCnt \leftarrow w.nodeCnt+1$\;
                    \If{$PQ$.\contain{$w$}}{
                        $PQ$.\remove{$w$}\;
                    }
                    $PQ$.\push{$w$}\;
                }
            }
            \Else{
                \update{$M[v]$, $N^g(v)$}\;
            }
        }
    }
    $VC$.\clear{}\;
    \While{$PQ$.\emptyc{}$\neq$ \true}{
        $v \leftarrow PQ$.\pop{}\; 
        \If{\getNodeLB{$v$, $M[v]$} $< k$}{
            \setcounter{AlgoLine}{36}
            \BlockIf{
                \setcounter{AlgoLine}{50}
            }
        }
    }
}
\Return{$V'$}
\caption{A procedure of {\ASAP}}
\label{alg:asap_alg}
\end{algorithm}

\spara{Algorithmic procedure.} The {\ASAP} shares a similar methodology with the {\NPA}, but improves efficiency by incorporating a lazy update strategy to reduce the number of $g$-neighbours computations. Algorithm~\ref{alg:asap_alg} provides a pseudo-description of the {\ASAP}. During the hyperedge peeling step, the $edgeCnt$ for nodes belonging to the removed hyperedges is incremented (Line 13). In the node peeling step, if \getEdgeLB{} with $edgeCnt$ returns a value of at least $k$, the computation of $g$-neighbours is not required, since the possible minimum $g$-neighbours size is larger than or equal to $k$ (Line 18). Additionally, the $nodeCnt$ for $g$-neighbours of removed nodes is increased (Line 27). If \getNodeLB{} with $nodeCnt$ returns a value of at least $k$, the computation of $g$-neighbours is skipped again (Line 36). This process is repeated until no more hyperedges or nodes remain to be removed.

\spara{Theoretical analysis.} 
The time complexity of Algorithm~\ref{alg:asap_alg} is the same as that of the {\NPA}, specifically $O(t|V|^2 |E|)$. This is because computing $N^g(v)$ requires $O(|V||E|)$ time, and the total number of iterations in the algorithm is bounded by the smaller of the two values, $|V|$ and $|E|$. Although both {\NPA} and {\ASAP} have the same theoretical time complexity, practical evaluations reveal notable performance differences. Empirical observations, discussed in Section~\ref{sec:experiments}, indicate that the {\ASAP} significantly enhances efficiency. This improvement is due to the more effective handling of computational tasks, especially in reducing redundant calculations, thus optimising the overall execution time.

\begin{example}
Considering Figure~\ref{fig:toy_graph_hyp} with parameters $k=2$, $g=2$, and $p=0.7$, the peeling procedure is as follows: First, nodes $\{v_5, \cdots, v_{12}\}$ are identified as a result of the $(k,g)$-core, and supporting tables for these nodes are initially based on their respective $g$-neighbours sizes. Due to the fraction constraint, hyperedges $e_1$ and $e_6$ are removed, as illustrated in Figure~\ref{fig:fraction_computation}. Consequently, the $edgeCnt$ for $\{v_5, \cdots, v_{10}\}$ is updated from $0$ to $1$, and for $\{v_{11}, v_{12}\}$, the $edgeCnt$ increases from $0$ to $2$. All nodes affected by $e_1$ and $e_6$ are added to the candidate node set $VC$ for further evaluation. As the \getEdgeLB{} function returns a result of $0$ for all nodes in $VC$, the $N^g(v)$ needs to be computed, leading to the removal of nodes $v_5$, $v_{11}$, and $v_{12}$. Consequently, hyperedges $e_2$ and $e_5$, which include these nodes, are added to the hyperedge candidate set $EC$. While $v_5$ and $v_{11}$ have no $g$-neighbours, $v_{12}$ maintains a $g$-neighbours relationship with $v_{10}$. Hence, the $nodeCnt$ for $v_{10}$ is increased by $1$, and $v_{10}$ is enqueued into the priority queue $PQ$. Subsequently, the \getNodeLB{} function is executed for $v_{10}$ in $PQ$, returning a value greater than $k$, and thus $v_{10}$ is retained. This process is repeated until {\ASAP} finally returns $\{v_6, \cdots, v_9\}$.
\end{example}

\section{REUSE STRATEGY FOR ITERATIVE VARYING $p$ }\label{sec:select_p}
Selecting an appropriate threshold parameter $p$ is crucial for the effective identification of cohesive subgraphs in hypergraphs. However, the optimal value of $p$ is often elusive, particularly in the absence of domain-specific background knowledge. Consequently, users may need to iteratively adjust $p$ and recompute the $(k,g,p)$-core to retrieve meaningful cohesive subhypergraphs. This section discusses the implications of varying $p$ and proposes an efficient strategy for reusing previously computed $(k,g,p)$-core.

\begin{figure}[t]
\begin{subfigure}{.99\linewidth}
\includegraphics[width=0.99\linewidth]{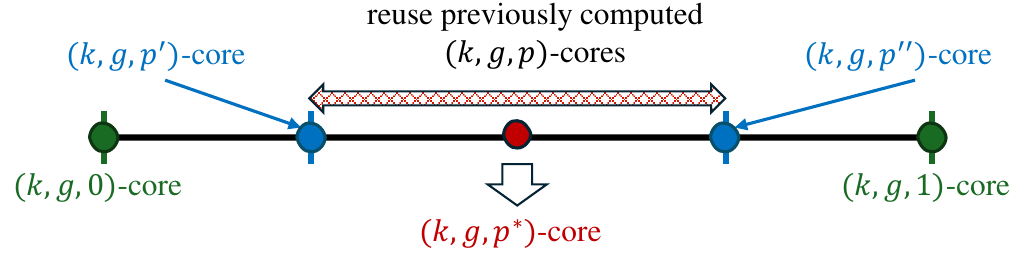}
\caption{A high-level overview of the reuse procedure}
\label{fig:reuse}
\end{subfigure}
\begin{subfigure}{.99\linewidth}
\vspace{0.2cm}
\includegraphics[width=0.99\linewidth]{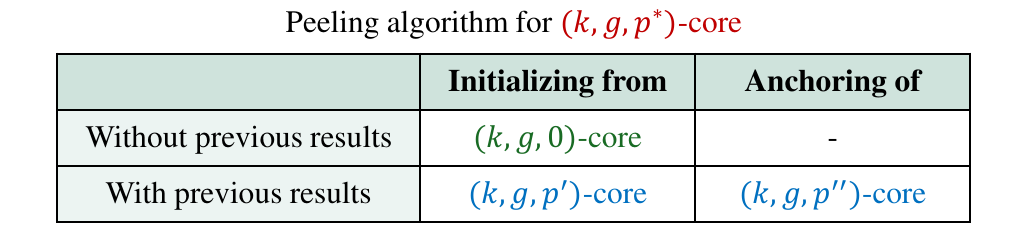}
\caption{Setting peeling algorithm from previous results}
\label{fig:reuse_table}
\end{subfigure}
\caption{Reuse strategy}
\label{fig:reuse_strategy}
\end{figure}

\spara{Reuse strategy.} To reduce the computational overhead of repeated $(k,g,p)$-core computations for different $p$ values, we propose a straightforward but effective reuse strategy. The high-level idea is to utilise previously computed $(k,g,p)$-cores to avoid redundant computations in the peeling algorithm. The reuse procedure is specified as follows.

Assume we have computed a series of $(k,g,p)$-cores with varying $p$ values. The main idea is to use these previous results to efficiently compute a new $(k,g,p^*)$-core. 
Figure~\ref{fig:reuse_strategy} illustrates a high-level overview of the reuse procedure. The process involves two key operations: (1) \textit{Efficient Initialisation with Previous Results:} For a given threshold $p^*$, identify the largest previously computed threshold $p'$ such that $p' < p^*$. This step helps to find the smallest superset as an initial point for the peeling procedure to compute the $(k,g,p^*)$-core. This allows us to start the computation from a close approximation, reducing the number of iterations required. (2) \textit{Setting Anchoring Nodes from Previous Results:} For the same threshold $p^*$, identify the smallest previously computed threshold $p''$ such that $p^* < p''$. This step helps to find the largest subset of nodes that should not be removed in the peeling process, as they are guaranteed to be part of the $(k,g,p^*)$-core.

\section{EXPERIMENTS}
\label{sec:experiments}
In this section, we evaluate the performance of our proposed algorithms on various real-world and synthetic datasets. Our goal is to demonstrate the effectiveness and efficiency of these algorithms.

\subsection{Experiment setup} 
To validate the superiority of our techniques, we conduct an extensive experimental study. The main objective is to get insights into the practical implications of our theoretical constructs and to evaluate them in diverse scenarios. Through our experiments, we aim to provide a clear understanding of the evaluation questions (\textbf{EQ}):
\begin{itemize}[leftmargin=*]
\item \textbf{EQ1. Effect on pruning strategies}: Since computing $g$-neighbours is a key task in {\ASAP}, this \textbf{EQ} evaluates the frequency of $g$-neighbours computations to check the impact on the pruning strategies, and to compare the effects on node-based and edge-based approaches.
\item \textbf{EQ2. Convergence analysis}: This \textbf{EQ} checks the number of nodes and hyperedges removed during each iteration to verify whether the algorithm converges within a few iterations.
\item \textbf{EQ3. Scalability}: This \textbf{EQ} evaluates the scalability of the proposed algorithms by varying the size of hypergraphs to evaluate the efficiency and performance of the algorithms.
\item \textbf{EQ4. Effect on user parameters}: This \textbf{EQ} evaluates the size of the resultant graph and the running time of the algorithms as the user parameters are varied.
\item \textbf{EQ5. Evaluation of reusing strategies for efficiency}: This \textbf{EQ} evaluates the efficiency of using previous results to obtain the $(k,g,p)$-core when the parameter $p$ is changed. The goal is to assess how effectively this approach reduces computational time.
\item \textbf{EQ6. Comparative analysis of $(k,g,p)$-core with existing models}: This \textbf{EQ} aims to compare the $(k,g,p)$-core with fraction-based and hypergraph-based models to observe trends in the resultant cohesive subgraphs and emphasising effectiveness of the {\kgp} model in capturing higher-order structures.
\item \textbf{EQ7. Case study}: This \textbf{EQ} evaluates the utility of the $(k,g,p)$-core in market data, which shows its effectiveness as a method for identifying co-purchased items and addressing the limitations of the $(k,g)$-core.
\end{itemize}

These evaluation questions guide our experimental analysis, enabling us to verify key aspects of the proposed pruning strategies and algorithmic approaches. The results from these experiments provide valuable insights into the performance and scalability of our methods from various perspectives, enhancing our understanding of their effectiveness and efficiency.

\subsection{Experimental setting}\label{sec:exp_setting}

\begin{table}[t]
\small
\caption{Real-world datasets}
\label{tab:data}
\centering
\begin{tabular}{c||c|c|c}
\hline
\textbf{Dataset} & $\boldsymbol{|V|}$ & $\boldsymbol{|E|}$ & $\boldsymbol{|E|}/\boldsymbol{|V|}$  \\ \hline \hline
Amazon~\cite{ni2019justifying} & $2,268,231$ & $4,285,363$ & $1.8893$ \\ \hline
AMiner~\cite{nbrkcore} & $27,850,748$ & $17,120,546$ & $0.6147$  \\ \hline
Gowalla~\cite{cho2011friendship} & $107,092$ & $1,280,969$ & $11.9614$  \\ \hline
House Bills~\cite{chodrow2021hypergraph, Fowler-2006-connecting,Fowler-2006-cosponsorship} & $1,494$ & $60,987$ & $40.8213$  \\ \hline
Instacart~\cite{Benson2018subset} & $49,677$ & $3,346,083$ & $67.3568$  \\ \hline
Kosarak~\cite{Benson2018subset} & $41,270$ & $990,002$ & $23.9884$  \\ \hline
\hline
\end{tabular}
\end{table}

\spara{Real-world dataset.} 
Table~\ref{tab:data} provides the detailed statistics of real-world datasets used in our experiments. 
These datasets are publicly available, as referenced~\cite{nbrkcore,chodrow2021hypergraph,Fowler-2006-connecting,Fowler-2006-cosponsorship,Benson2018subset,cho2011friendship,ni2019justifying}. We transformed the location-based social network (Gowalla dataset) into a hypergraph representation by utilising each location ID as a hyperedge and the user IDs of visitors to that location serving as the nodes of the hyperedge.

\spara{Parameter setting.}
Since the degree or hyperedge cardinality distribution varies significantly across datasets, we set different default parameters and their variations for two distinct groups of datasets based on their size. The specific values and variations of these parameters are detailed in Table~\ref{tab:exp_param_variation}. The default values are underlined. 

\begin{table}[h]
\small
\caption{\mbox{Parameter variations}}
\label{tab:exp_param_variation}
\centering
\begin{tabular}{c||c|c|c}
\hline
\textbf{Dataset} & \textbf{Var} & \textbf{Variation \{$\boldsymbol{x_{1}-x_{5}}$\}} & \textbf{Threshold for} \\ \hline \hline
\multirow{3}{*}{\begin{tabular}[c]{@{}c@{}}Amazon,\\ AMiner,\\ Gowalla \end{tabular}} 
 & $k$ & $3, 4, \underline{5}, 6, 7$ & $g$-neighbours size \\ \cline{2-4} 
 & $g$ & $3, 4, \underline{5}, 6, 7$ & support \\ \cline{2-4} 
 & $p$ & $0.2, 0.4, 0.6, \underline{0.8}, 1.0$ & fraction  \\ \hline
\multirow{3}{*}{\begin{tabular}[c]{@{}c@{}}House Bills,\\ Instacart,\\ Kosarak \end{tabular}}
 & $k$ & $20, 25, \underline{30}, 35, 40$ & $g$-neighbours size \\ \cline{2-4} 
 & $g$ & $20, 25, \underline{30}, 35, 40$ & support \\ \cline{2-4} 
 & $p$ & $0.2, 0.4, 0.6, \underline{0.8}, 1.0$ & fraction \\ \hline \hline
\end{tabular}
\end{table}

\spara{Algorithms.} 
To the best of our knowledge, our problem does not have any direct competitors. Fraction-based approaches~\cite{alphabetap,kpcore} are applicable to unipartite and bipartite graphs. Although \cite{bu2023hypercore} is designed for hypergraphs, it does not consider the support value of the neighbour structure, so direct comparison is challenging. Thus, we compare the proposed algorithms with the $(k,g)$-core (Section~\ref{sec:kgcore_comp}) in our main experiments. For some experimental tasks, we use other fraction-based approaches to check the results when the fraction threshold increases. 

\spara{Experimental environment.} All experiments were run on a Linux machine with Intel Xeon $6248$R and $256$GB of RAM.

\subsection{Experimental result}\label{sec:exp_result}

\begin{figure}[h]
    \includegraphics[width=0.99\linewidth]{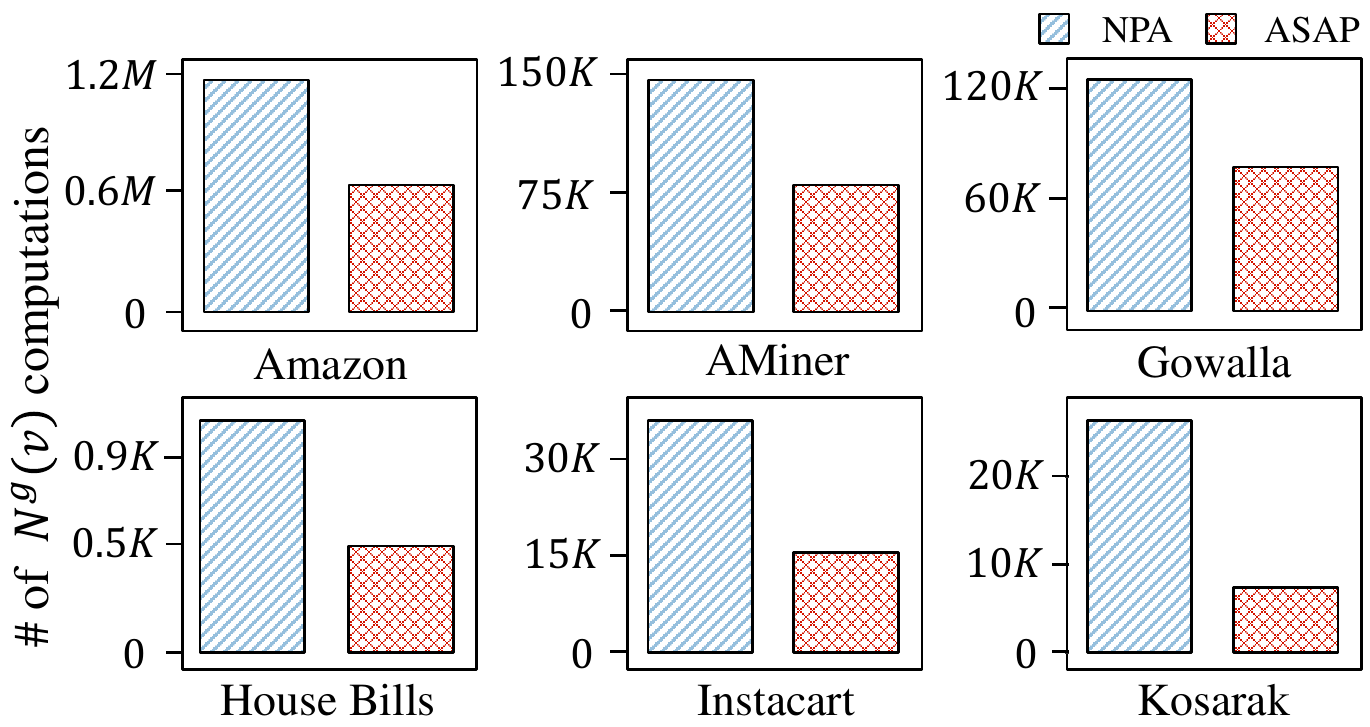}
    \caption{\textbf{EQ1-1.} The number of $N^g(v)$ operations}
    \vspace{-0.2cm}
    \label{fig:case1}
\end{figure}

\spara{EQ1-1. Effect on pruning strategies (\# of $g$-neighbours computation).}
In our comparative analysis of {\NPA} and {\ASAP}, we focused on the frequency of $N^g(v)$ operations required by each method, as depicted in Figure~\ref{fig:case1}. The results clearly demonstrate that the {\ASAP} significantly reduces the number of $N^g(v)$ operations across all datasets by over $40\%$, effectively halving the computations needed compared to the {\NPA}. Even in the Gowalla dataset, which showed the lowest reduction, the {\ASAP} performed approximately $37.9\%$ fewer calculations. In contrast, the Kosarak dataset exhibited a marked improvement, with approximately $72\%$ fewer calculations. These results highlight the effectiveness of the {\ASAP}, making it highly advantageous for applications requiring frequent $N^g(v)$ computations. By reducing these operations, {\ASAP} not only accelerates computation but also lessens the computational load.

\begin{figure}[h]
    \includegraphics[width=0.99\linewidth]{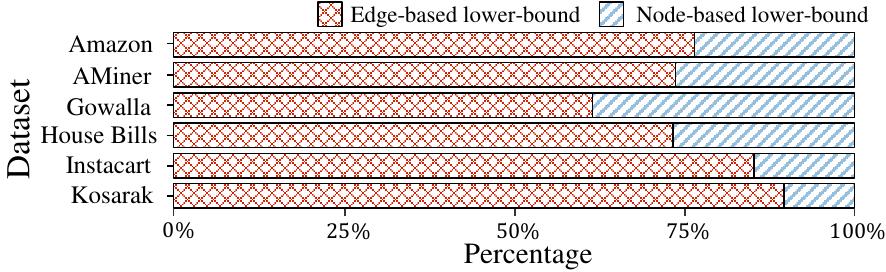}
    \caption{\mbox{\textbf{EQ1-2.} Contributions of node/edge-based lower-bounds}}
    \vspace{-0.2cm}
    \label{fig:case1_2}
\end{figure}

\spara{EQ1-2. Effect on node-based and edge-based lower-bounds.}
We analyse the impact of our proposed lazy update strategy, which effectively avoids the computationally expensive task of calculating $g$-neighbours. Figure~\ref{fig:case1_2} illustrates the contributions of both node-based and edge-based lower-bounds in avoiding $g$-neighbours computations. Overall, the edge-based lower-bounds account for an average of $76.6\%$ in lazy updates, significantly surpassing the node-based lower-bounds. In the Kosarak dataset, edge-based lower-bounds contribute $89.6\%$, while in the Gowalla dataset, they contribute a minimum of $61.5\%$. We observed that edge-based lower-bounds are instrumentally important in optimising the efficiency of {\ASAP} by reducing the frequency of $g$-neighbours calculations.

\begin{figure}[h]
\centering
\begin{subfigure}{.99\linewidth}
\includegraphics[width=0.99\linewidth]{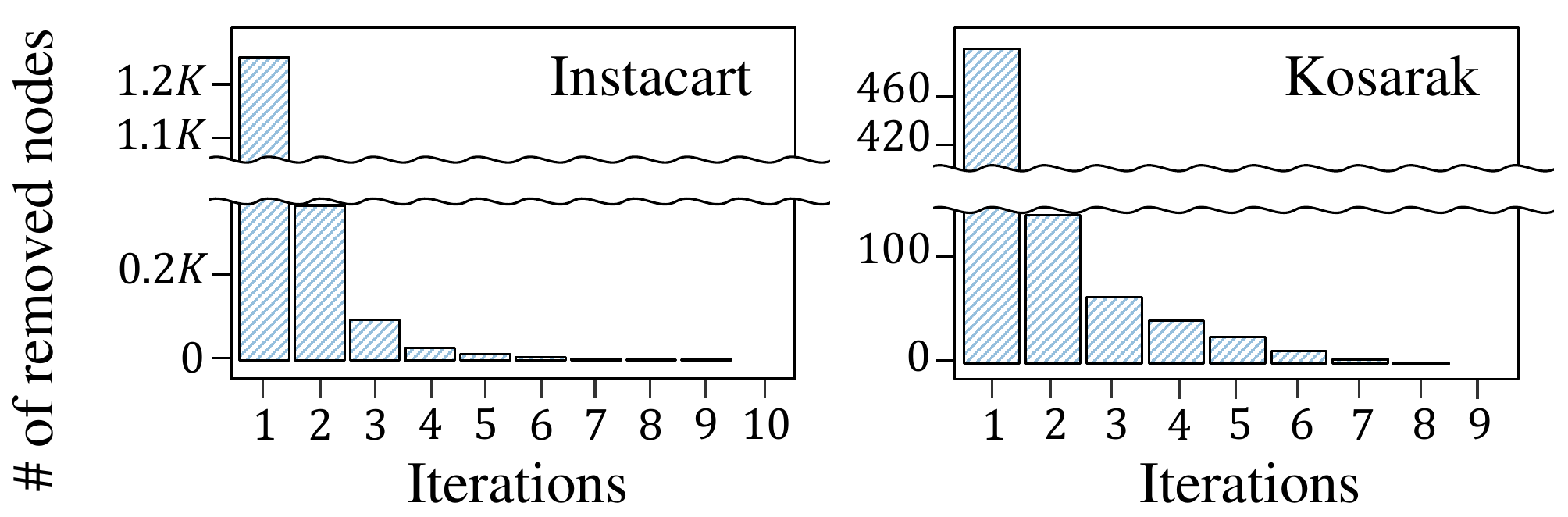}
\caption{\textbf{Number of nodes removed per iteration}}
\label{fig:case3_node}
\end{subfigure}
\begin{subfigure}{.99\linewidth}
\includegraphics[width=0.99\linewidth]{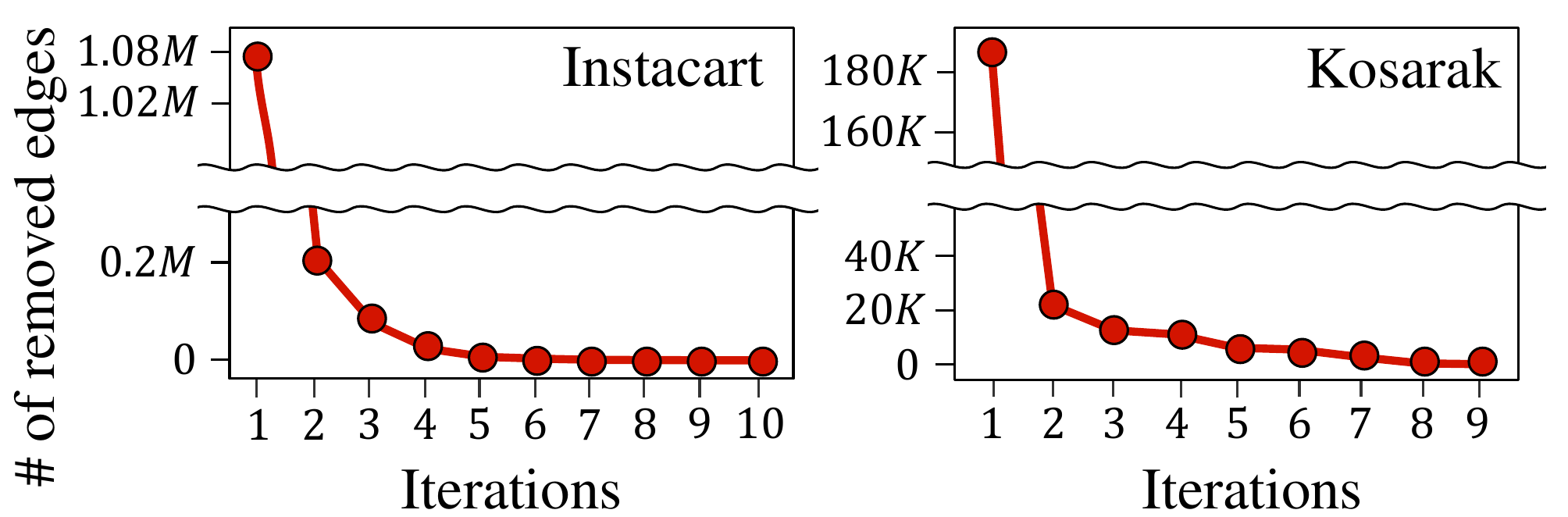}
\caption{\textbf{Number of hyperedges removed per iteration}}
\label{fig:case3_edge}
\end{subfigure}
\vspace{-0.2cm}
\caption{\textbf{EQ2}. Convergence analysis}
\label{fig:case3}
\end{figure}

\spara{\textbf{EQ2.} Convergence analysis.}
We analyse the efficiency of our peeling algorithms by checking the number of nodes and hyperedges removed at each iteration on various datasets. We examined the number of iterations, observing rapid convergence within a few iterations, although the maximum number of iterations is bounded by $\min(|V|, |E|)$. The average number of iterations was $14.6$. Interestingly, the House Bills dataset required the fewest iterations, with only $3$, whereas the Amazon dataset needed the most, at $39$ iterations. In the initial iterations (up to $2$-$3$ iterations), we observed that over $85\%$ of the total eliminated nodes and hyperedges were removed, indicating rapid convergence towards the end result. Figure~\ref{fig:case3} presents this tendency, showing that significant reductions are achieved with just a few iterations in both cases.

\begin{figure}[h]
    \includegraphics[width=0.99\linewidth]{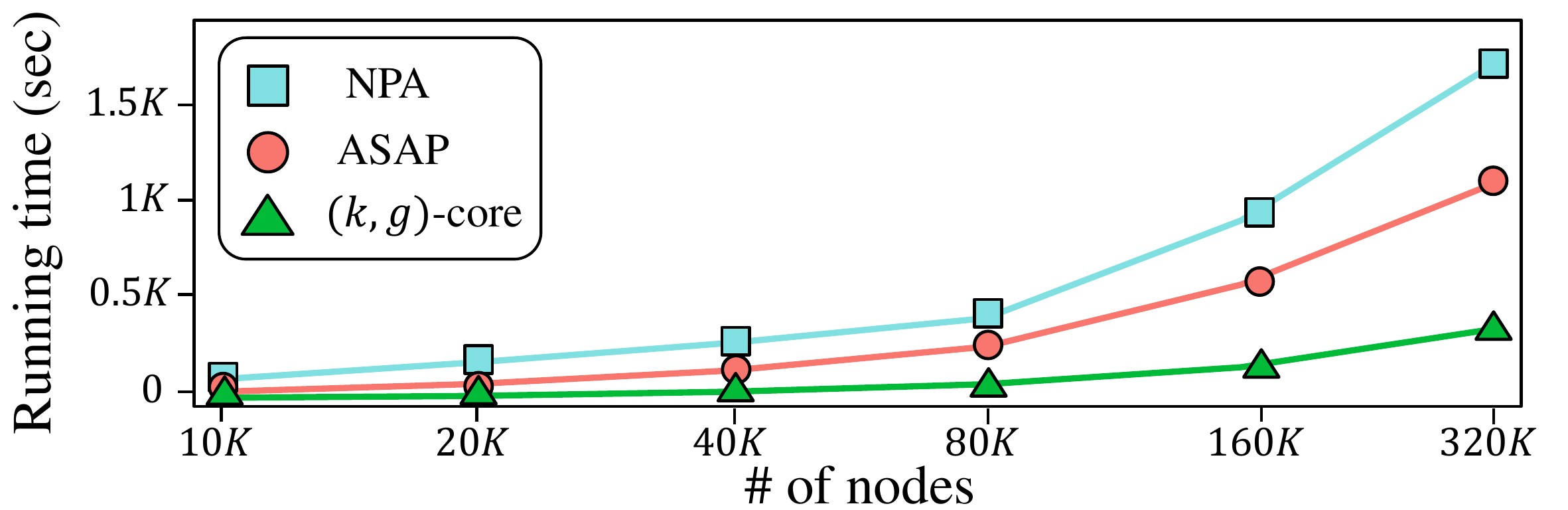}
    \vspace{-0.2cm}
    \caption{\textbf{EQ3.} Scalability test}
    \label{fig:scalability}
    \vspace{-0.2cm}
\end{figure}

\spara{EQ3. Scalability.} Figure~\ref{fig:scalability} illustrates the scalability of our two algorithms and the $(k,g)$-core computation algorithm (Section~\ref{sec:kgcore_comp}) when we vary the hypergraph node size from $10,000$ to $320,000$. The synthetic hypergraphs are generated by the H-ABCD model~\cite{kaminski2023hypergraph} with the following parameter setting: node degrees are distributed according to a heavy-tailed distribution with a parameter of $2.1$, ranging from $40$ to $1,000$. Hyperedges are uniformly distributed, with sizes varying from $1$ to $40$, and include a noise level of $0.2$. In this experiment, we use the default parameters for $k$, $g$, and $p$ as specified in Section~\ref{sec:exp_setting}. We observe that {\ASAP} significantly outperforms the existing {\NPA} in terms of running time, which is due to the avoidance of repetitive $g$-neighbours computations. With $320,000$ nodes, {\ASAP} running time is $35.6\%$ faster than {\NPA}.

\begin{figure}[!b]
\centering
\begin{subfigure}{.99\linewidth}
\includegraphics[width=0.99\linewidth]{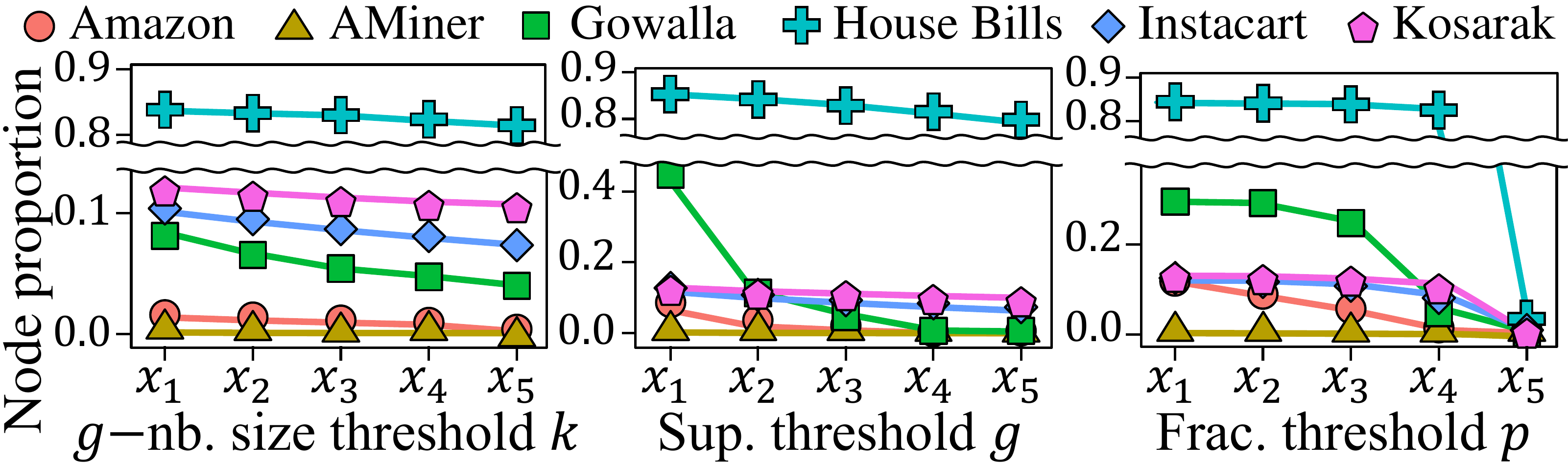}
\caption{\textbf{Remained node size of the {\kgp}}}
\vspace{0.2cm}
\label{fig:case2_1_node}
\end{subfigure}
\begin{subfigure}{.99\linewidth}
\includegraphics[width=0.99\linewidth]{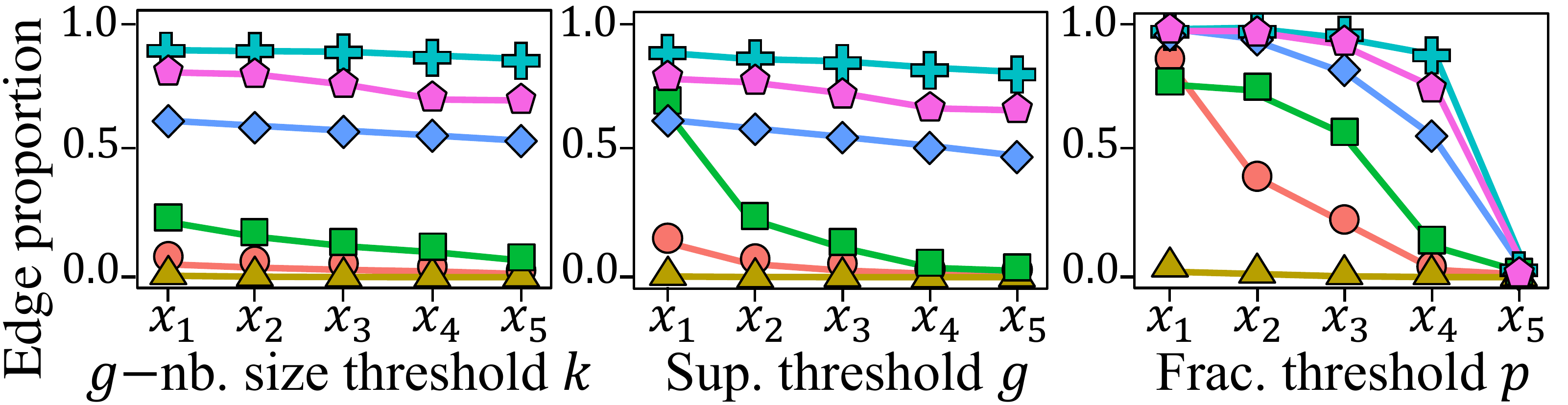}
\caption{\textbf{Remained hyperedge size of the {\kgp}}}
\vspace{0.2cm}
\label{fig:case2_1_edge}
\end{subfigure}
\caption{\textbf{EQ4-1}. Effect on various parameter settings}
\label{fig:case2_1}
\vspace{-0.2cm}
\end{figure}

\spara{EQ4-1. Effect on user parameters (size).}
The effect of the {\kgp} by varying user parameters $k$, $g$, and $p$ is evaluated by checking the number of resultant nodes and hyperedges, as shown in Figure~\ref{fig:case2_1}. In all cases, when the values of $k$, $g$, and $p$ are increased, the proportion of nodes and hyperedges consistently decreases. The House Bills dataset has the highest node and hyperedge proportion, slightly above $0.8$, while the minimum is less than $0.01$ in the AMiner dataset. Overall, varying $k$ and $g$ results in a relatively linear decrease in resultant nodes and hyperedges. However, the parameter $p$ introduces points of rapid convergence, indicating that the performance of the {\kgp} is less stable with respect to $p$. An interesting observation is that a large number of nodes and hyperedges seems correlated with the effectiveness of pruning strategies (Figure~\ref{fig:case1}) and a high proportion of edge-based lower-bounds (Figure~\ref{fig:case1_2}).

\begin{figure}[h]
\centering
\begin{subfigure}{.99\linewidth}
\includegraphics[width=0.99\linewidth]{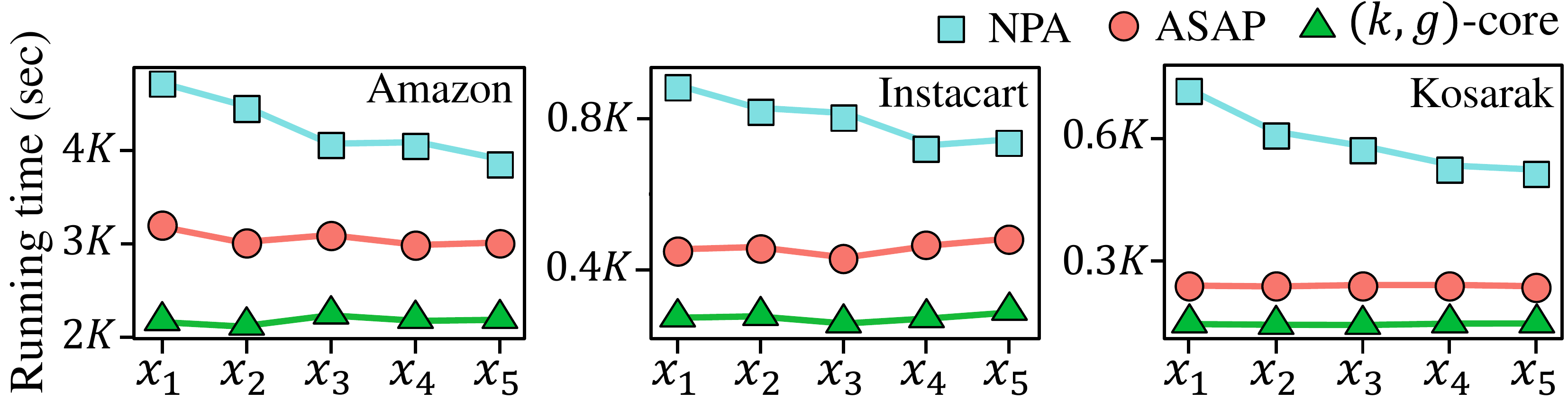}
\caption{\textbf{Varying $g$-neighbours size threshold $k$}}
\vspace{0.2cm}
\label{fig:case2_2k}
\end{subfigure}
\begin{subfigure}{.99\linewidth}
\includegraphics[width=0.99\linewidth]{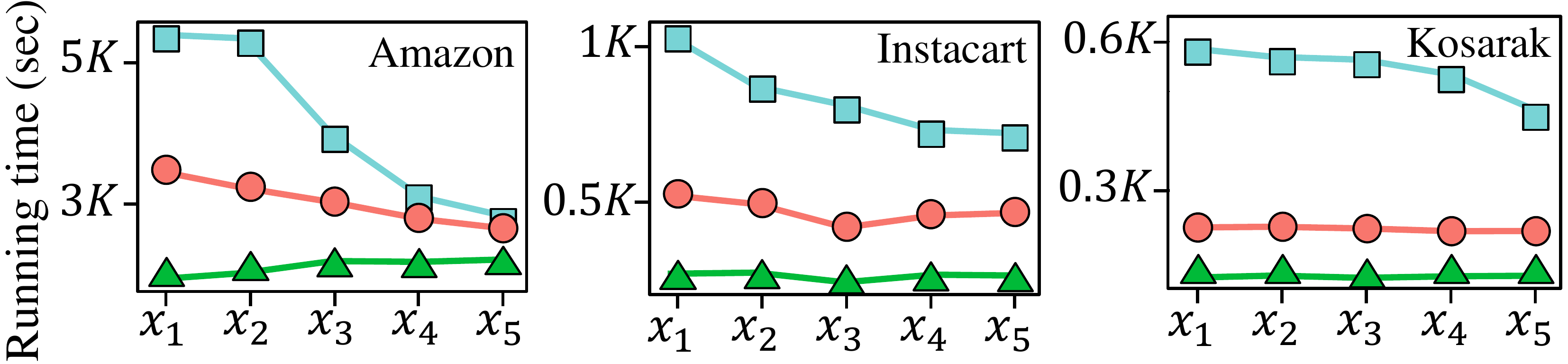}
\caption{\textbf{Varying support threshold $g$}}
\vspace{0.2cm}
\label{fig:case2_2g}
\end{subfigure}
\begin{subfigure}{.99\linewidth}
\includegraphics[width=0.99\linewidth]{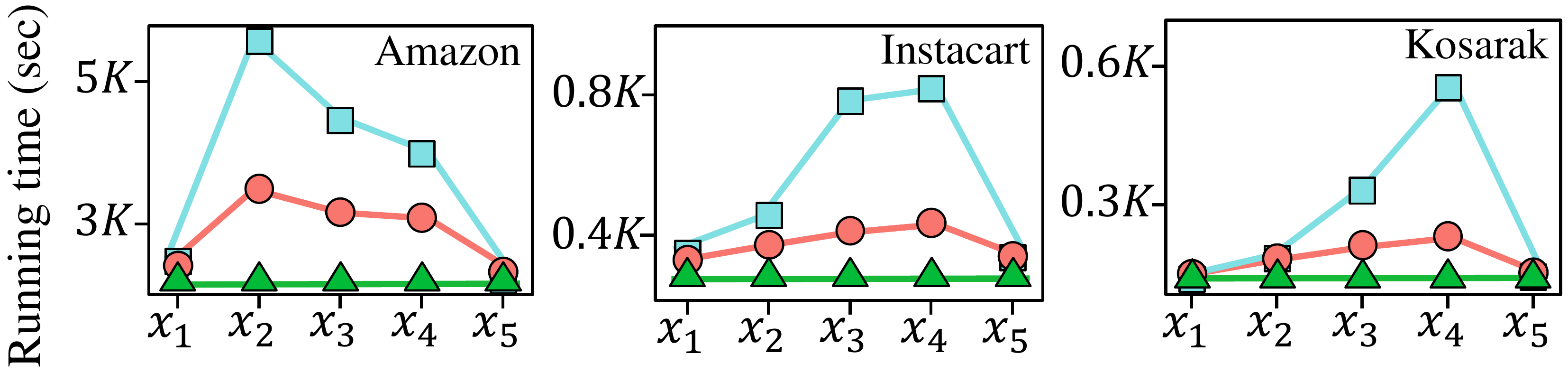}
\caption{\textbf{Varying fraction threshold $p$}}
\label{fig:case2_2p}
\end{subfigure}
\caption{\textbf{EQ4-2}. Comparison of running time on parameters}
\label{fig:case2_2}
\vspace{-0.2cm}
\end{figure}

\spara{EQ4-2. Effect on user parameters (running time). }
We compare the running times of our algorithms and the $(k,g)$-core across various user parameters $k,g$ and $p$. Figure~\ref{fig:case2_2} demonstrates that the running time of the {\ASAP} is on average $1.7$ times faster than that of the {\NPA}, and the $(k,g)$-core computation takes about $65\%$ of the total running time of the {\ASAP}. When varying $k$ and $g$, the running time generally decreases linearly. However, with variations in parameter $p$, the {\NPA} exhibits an irregular pattern. During increased running time, the actual number of iterations varies significantly: from $8$ to $33$ in the Amazon dataset, from $2$ to $10$ in the Instacart dataset, and from $3$ to $9$ in the Kosarak dataset. This indicates that hyperedges are removed in a cascading manner, requiring {\NPA} to compute $g$-neighbours multiple times, which increases the running time. 

\begin{figure}[t]
    \includegraphics[width=0.99\linewidth]{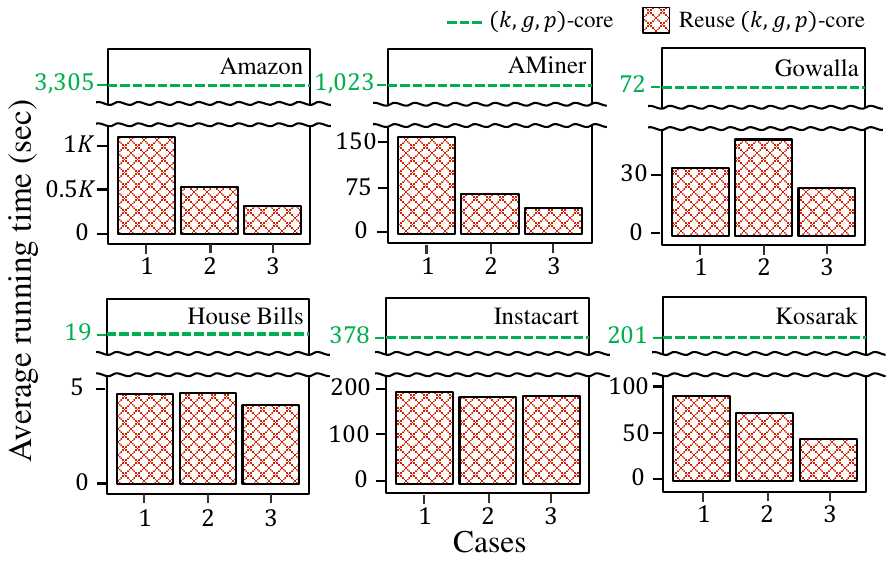}
    \vspace{-0.2cm}
    \caption{\textbf{EQ5.} Comparison of running time on cases }
    \vspace{-0.2cm}
    \label{fig:case5}
\end{figure}

\spara{EQ5. Evaluation of reusing strategies for efficiency.}
To evaluate the effectiveness of reusing strategies, we considered three different scenarios. The settings are detailed in Table~\ref{tab:threshold}. We have 3 different cases based on the search space. Figure~\ref{fig:case5} shows the running times for two categories: (1) average running time without using previous results (green dotted line) and (2) average running time with utilising previous results (red bar). We observed that, in all cases, the running time improves when previous results are utilised. Also, in most cases, the running time is decreased when we use more previous results. However, in the Gowalla dataset, even though previous results were used in cases \#2, efficiency is worse than case \#1. We found that in case \#2-2, the number of nodes was reduced by approximately $75\%$ compared to case \#1. This suggests that significant node and hyperedge removal occurs when $p$ is greater than $0.5$.

\begin{table}[t]
\centering
\caption{Case scenarios}
\label{tab:threshold}
\begin{tabular}{c||c|c|c}
\hline
\textbf{Case} & \textbf{Label} & \textbf{Threshold $\boldsymbol{p}$} & \textbf{Previous results} \\ \hline \hline
$0$ & case \#$0$ & $1$ & NA \\ \hline
$1$ & case \#$1$ & $0.5$ & case \#$0$ \\ \hline
\multirow{2}{*}{$2$} & case \#$2$-$1$ & $0.25$  & \multirow{2}{*}{case \#$0$ $\sim$\#$1$} \\ \cline{2-3}
                     & case \#$2$-$2$ & $0.75$  & \\ \hline
\multirow{4}{*}{$3$} & case \#$3$-$1$ & $0.125$ & \multirow{4}{*}{case \#$0$ $\sim$case \#$2$} \\ \cline{2-3}
                     & case \#$3$-$2$ & $0.375$ & \\ \cline{2-3}
                     & case \#$3$-$3$ & $0.625$ & \\ \cline{2-3}
                     & case \#$3$-$4$ & $0.875$ & \\ \hline \hline
\end{tabular}
\end{table}

\spara{EQ6-1. Comparative analysis of $(k,g,p)$-core with fraction-based models. } 
We compare the performance of our $(k,g,p)$-core with other fraction-based models, including $(k,t)$-hypercore~\cite{bu2023hypercore}, $(k,p)$-core~\cite{kpcore} and $(\alpha, \beta, p)$-core~\cite{alphabetap}, to analyse trends in the fraction-based cohesive subgraphs. Given that the input graph types differ for $(k,p)$-core and $(\alpha,\beta,p)$-core, we apply the following transformation procedures: (1) For the $(k,p)$-core, we convert the hypergraph into a clique-based graph, meaning that each hyperedge forms a clique; (2) For the $(\alpha,\beta,p)$-core, we consider the set of nodes in the hypergraph as one part and the set of hyperedges as the other. If a node belongs to a hyperedge, we create an edge between the corresponding nodes in the bipartite graph.

\begin{figure}[t]
    \includegraphics[width=0.99\linewidth]{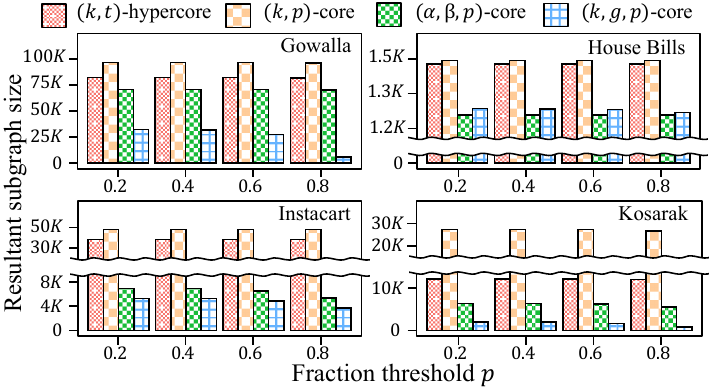}
    \vspace{-0.1cm}
    \caption{\textbf{EQ6-1}. Comparison with fraction-based models}
    \vspace{-0.2cm}
    \label{fig:case6}
\end{figure}

For our experiments, we use the default parameters specified in Table~\ref{tab:exp_param_variation} for the $(k,g,p)$-core. For the $(k,t)$-hypercore and the $(k,p)$-core, we apply the same $k$ value, and for the $(\alpha,\beta,p)$-core, we set $\alpha = k$ and $\beta = k$. The experimental results are presented in Figure~\ref{fig:case6}. Our observations indicate that the {\kgp} demonstrates stronger cohesiveness in hypergraphs compared to other fraction-based model. This is due to its stricter support and fraction threshold compared to the degree or cardinality constraints. Notably, the $(k,g,p)$-core demonstrates a more responsive changes in the fraction threshold $p$. When $p$ increases, the size of the cohesive subgraph consistently decreases compared to other fraction-based models, demonstrating the sensitivity of the $(k,g,p)$-core to this parameter. This observation emphasises the importance of the fraction threshold in determining the granularity of the cohesive subgraphs detected by $(k,g,p)$-core.

\begin{figure}[t]
    \includegraphics[width=0.99\linewidth]{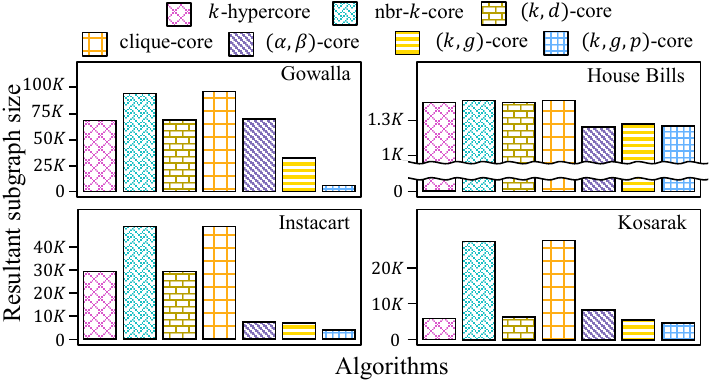}
    \vspace{-0.1cm}
    \caption{\textbf{EQ6-2}. Comparison with hypergraph-based models}
    \vspace{-0.3cm}
    \label{fig:case6_2}
\end{figure}

\begin{figure*}[t]
    \centering
    \includegraphics[width=0.9\linewidth]{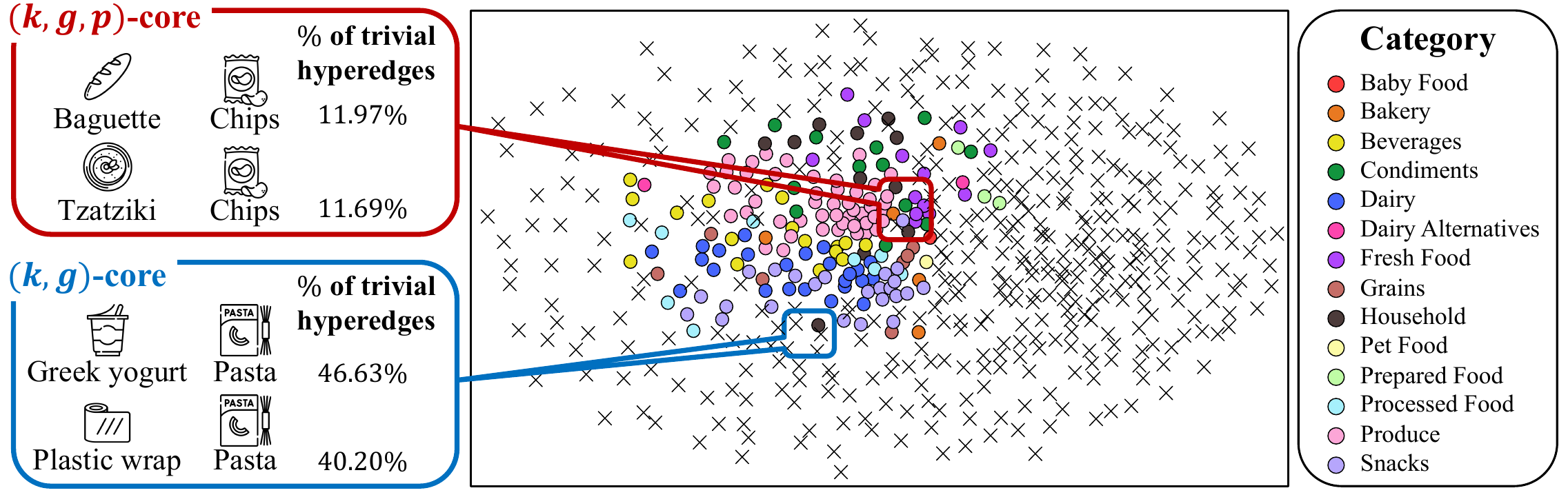}
    \vspace{-0.2cm} 
    \caption{\textbf{EQ7}. Case study on the Instacart dataset illustrating the {\kgp} with $k=100$, $g=150$, and $p=0.5$. Only nodes are shown for clarity, with those in the {\kgp} coloured by product categories and cross-marked nodes representing those in the {\kg} but removed during the {\kgp} pruning process. The left panel compares node pairs with the highest proportion of trivial hyperedges, comparing in the {\kg} (blue) and {\kgp} (red).}
    \label{fig:case7}
    \vspace{-0.1cm}
\end{figure*}

\spara{EQ6-2. Comparative analysis of $(k,g,p)$-core with hypergraph-based models.}
We compare the performance of our {\kgp} with other hypergraph-based cohesive subgraph models, including $k$-hypercore~\cite{leng2013m}, nbr-$k$-core~\cite{nbrkcore}, $(k,d)$-core~\cite{nbrkcore}, Clique-core~\cite{batagelj2011fast}, $(\alpha,\beta)$-core~\cite{alphabeta}, and $(k,g)$-core~\cite{kgcore}. For Clique-core and $(\alpha,\beta)$-core, we apply the same transformation methods used in EQ6-1. We use the default parameter $k$ for all models, as specified in Table~\ref{tab:exp_param_variation}. In the case of $(k,d)$-core, we set $k=d$, and for $(\alpha,\beta)$-core, we set $\alpha=\beta=k$. For $(k,g)$-core and our $(k,g,p)$-core, we apply the default values.

The experimental results are presented in Figure~\ref{fig:case6_2}. Across all four datasets, the average number of neighbours is extremely high, ranging from $664.10$ in the House Bills dataset to $1,805.73$ in the Instacart dataset. As a result, neighbour-based models such as nbr-$k$-core and Clique-core easily satisfy the neighbour requirement and thus undergo almost no pruning. In contrast, models that utilise node degrees and hyperedge cardinalities---$k$-hypercore, $(k,d)$-core, and $(\alpha,\beta)$-core—apply moderately stronger constraints and therefore produce subgraphs of intermediate size. Compared to these na\"ive constraints, the $(k,g)$-core enforces a stricter condition requiring each node to repeatedly co-occur with multiple neighbours across several shared hyperedges, resulting in smaller subgraphs. The $(k,g,p)$-core further tightens this condition by incorporating the hyperedge fraction, yielding the smallest subgraphs among all models. These results demonstrate that the $(k,g,p)$-core effectively captures cohesive structures in hypergraphs by integrating both support-based and fraction-based constraints.

\spara{EQ7. Case study.} 
Figure~\ref{fig:case7} illustrates the effectiveness of the {\kgp} on the Instacart dataset. Approximately $90\%$ of hyperedges consist of $20$ or fewer items, representing typical shopping patterns. Consequently, we consider hyperedges with more than $20$ items as trivial hyperedges, as they might disproportionately influence the subhypergraph structure and dilute meaningful connectivity. For instance, "Greek yogurt and Pasta" and "Plastic wrap and Pasta" included in the {\kg}, contain up to $46\%$ of their total hyperedges as trivial hyperedges, which weakens meaningful relationships and complicates the identification of cohesive structures. As a result, the pruning process excludes these pairs from the {\kgp}. In contrast, in the {\kgp}, it contains "Baguette", "Chips", and "Tzatziki" as a result of {\kgp}. Note that "Baguette and Chips" and "Tzatziki and Chips" retain only $11\%$ of their hyperedges as trivial hyperedges. This proportion is at least $3.43$ times lower than in {\kg}, emphasising that the {\kgp} is relatively more effective than the {\kg} at reducing trivial hyperedges. Consequently, the {\kgp} facilitates the discovery of more cohesive and meaningful substructures.

\section{RELATED WORK} \label{sec:relatedwork}
In this section, we discuss several related works, including cohesive subgraph discovery in hypergraphs and fraction-based models in various networks. Table~\ref{tab:relatedworks} summarises existing models alongside our proposed model.

\subsection{Cohesive subgraph models in hypergraphs}
This section explores various cohesive subgraph models in hypergraphs that address different aspects of complex relationships. The $k$-hypercore~\cite{leng2013m}, as the first extension of the $k$-core~\cite{seidman1983network} for hypergraphs, ensures that each node has a minimum degree but lacks the ability to capture high-order relationships due to its limited constraint. To overcome this limitation, the $(k,t)$-hypercore~\cite{bu2023hypercore} introduces a proportional threshold for hyperedge inclusion. However, it does not account for the distinct roles or contributions of individual nodes within the same hyperedges, thereby limiting its ability to capture complex, high-order relationships among nodes. The nbr-$k$-core~\cite{nbrkcore} introduces a ``strongly induced subhypergraph'' where a hyperedge exists if and only if all its nodes are within that subhypergraph. This model addresses the limitations of the $k$-hypercore by considering the neighbourhood structure, but it still struggles with large hyperedges and strict constraints. The $(k,d)$-core~\cite{nbrkcore} adds a degree constraint to mitigate the limitation of nbr-$k$-core, but the strict constraints may still exclude significant cohesive structures. The Clique-core~\cite{batagelj2011fast} transforms hypergraphs into graphs, identifying $k$-cores but potentially diluting the rich interconnectivity of hypergraphs. Lastly, the $(\alpha, \beta)$-core~\cite{alphabeta} models cohesive structures in bipartite graphs for dual relationship analysis, but both models suffer from size inflation when converting hypergraphs~\cite{huang2015scalable}. These models offer various approaches to understanding the structure of hypergraphs but often fall short in capturing the full complexity of relationships. Furthermore, recent advancements have enhanced cohesive subgraph discovery in hypergraphs. A locality-driven indexing framework has been proposed to efficiently retrieve cohesive subgraphs, optimising query processing and reducing computational costs~\cite{kim2025cohesive}. Additionally, a $k$-core decomposition method improves scalability in billion-scale hypergraphs by using core weights and pruning to reduce redundancy and memory usage~\cite{zhang2025accelerating}.

\subsection{Cohesive subgraph discovery with fractions}
This section introduces models that utilise edge fractions to define cohesive subgraphs. The $(\alpha, \beta, p)$-core~\cite{alphabetap} extends bipartite graph analysis by incorporating the fraction constraint, enhancing the detection of internally engaged cohesive subgraphs. However, like the $(\alpha, \beta)$-core, it suffers from the problem of size inflation~\cite{huang2015scalable}. Similarly, the $(k,p)$-core~\cite{kpcore} extends the traditional $k$-core~\cite{seidman1983network} by adding a fraction constraint to ensure denser connectivity within cohesive subgraphs in unipartite networks. Recently, a local search approach was proposed for efficiently maintaining $(k,p)$-core structures in dynamic graphs~\cite{zhang2025local}. However, it may still struggle to fully capture multi-layered relationships. By applying the fraction constraint, both models effectively reduce the influence of hub nodes and enhance the identification of cohesive structures in networks. In hypergraphs, the fraction constraint enables the exclusion of broad and trivial hyperedges, facilitating the discovery of more meaningful and cohesive subhypergraphs. These approaches demonstrate the effectiveness of fraction constraints in enhancing the quality of subgraph discovery.

\begin{table}[h]
\small
\centering
\caption{Comparison of models (\textbf{N}: neighbour size, \textbf{D}: degree, \textbf{C}: cardinality, \textbf{F}: edge fraction, \textbf{S}: support, \textbf{P} : allowance of partial participation from hyperedge)}
\label{tab:relatedworks}
\begin{tabular}{c||c|c|c|c|c|c|c} \hline
\multirow{2}{*}{\textbf{Model}}  & \multicolumn{6}{c|}
{\textbf{Constraints}}  & \multirow{2}{*}{\textbf{Type}} \\ \cline{2-7}
& \multicolumn{1}{c|}{N} & \multicolumn{1}{c|}{D} & \multicolumn{1}{c|}{C} & \multicolumn{1}{c|}{F}  & \multicolumn{1}{c|}{S}  & \multicolumn{1}{c|}{P}  \\ \hline \hline$k$-hypercore~\cite{leng2013m}               & $\bigtimes$  & $\bigcirc$  & $\bigtimes$  & $\bigtimes$ & $\bigtimes$ & $\bigtimes$  & Hyper \\ \hline
\hspace{-0.1cm}$(k,t)$-hypercore~\cite{bu2023hypercore}   \hspace{-0.2cm}  & $\bigtimes$ & $\bigcirc$  & $\bigtimes$  & $\bigcirc$ & $\bigtimes$  & $\bigcirc$  & Hyper \\ \hline
nbr-$k$-core~\cite{nbrkcore}                 & $\bigcirc$ & $\bigtimes$  & $\bigtimes$  & $\bigtimes$ & $\bigtimes$  & $\bigtimes$  & Hyper \\ \hline
$(k,d)$-core~\cite{nbrkcore}                 & $\bigcirc$ & $\bigcirc$  & $\bigtimes$  & $\bigtimes$ & $\bigtimes$  & $\bigtimes$  & Hyper \\ \hline
Clique-core~\cite{batagelj2011fast}          & $\bigcirc$ & $\bigtimes$  & $\bigtimes$  & $\bigtimes$ & $\bigtimes$  & $\bigcirc$  & \hspace{-0.15cm}Unipartite\hspace{-0.15cm}  \\ \hline
$(\alpha, \beta)$-core~\cite{alphabeta}      & $\bigtimes$ & $\bigcirc$  & $\bigcirc$ & $\bigtimes$  & $\bigtimes$  & $\bigcirc$  & Bipartite \\ \hline
$(k,g)$-core~\cite{kgcore}                   & $\bigcirc$ & $\bigtimes$  & $\bigtimes$ & $\bigtimes$  & $\bigcirc$  & $\bigcirc$  & Hyper \\ \hline
$(\alpha, \beta, p)$-core~\cite{alphabetap}  & $\bigtimes$ & $\bigcirc$  & $\bigcirc$ & $\bigcirc$  & $\bigtimes$  & $\bigcirc$  & Bipartite \\ \hline
$(k,p)$-core~\cite{kpcore}                   & $\bigcirc$ & $\bigtimes$  & $\bigtimes$ & $\bigcirc$  & $\bigtimes$  & $\bigcirc$  & \hspace{-0.15cm}Unipartite\hspace{-0.15cm} \\ \hline
\boxit{-19.5pt}{7.4cm}{2pt} $(k,g,p)$-core    & $\bigcirc$ & $\bigtimes$  & $\bigtimes$ & $\bigcirc$  & $\bigcirc$  & $\bigcirc$  & Hyper \\ \hline \hline
\end{tabular}
\end{table}

\section{CONCLUSION}
\label{sec:conclusion}
This paper introduced the $(k,g,p)$-core model, improving the $(k,g)$-core by incorporating a fraction constraint to better identify cohesive subhypergraphs. This model addresses the limitation of filtering out irrelevant hyperedges, resulting in more accurate subhypergraph identification. We also integrated node-based and edge-based pruning strategies with lazy updates, which reduces redundant computations and improves efficiency, making the $(k,g,p)$-core model suitable for large-scale applications. Future work involves adaptive thresholds that dynamically adjust based on hypergraph characteristics and designing advanced edge-based lower bounds to improve support value updates. Additionally, we plan to design more tight edge-based lower bounds to further improve efficiency.

\section*{ACKNOWLEDGMENT}
This work was supported by Institute of Information \& communications Technology Planning \& Evaluation(IITP) grant funded by the Korea government(MSIT)
(No.RS-2020-II201336, Artificial Intelligence Graduate School Program(UNIST)), the National Research Foundation of Korea(NRF) grant funded by the Korea government(MSIT) (RS-2023-00214065, RS-2023-00278009, RS-2025-00523578)
, Basic Science Research Program through the
National Research Foundation of Korea(NRF) funded by the Ministry of Education(No.RS-2024-00354951)
and Korea Foundation for Women In Science, Engineering and Technology (WISET) grant, funded by the Ministry of Science and ICT(MSIT) under the Team 
Research Program for female engineering students. (WISET No.2025-175)

\bibliographystyle{ACM-Reference-Format}
\bibliography{sample-base}

\end{document}